\documentclass[aps,prd,amsmath,floats,floatfix,twocolumn,superscriptaddress,nofootinbib,noshowpacs]{revtex4-2}
\usepackage{amssymb,amsmath,mathtools,dsfont,latexsym}
\usepackage{amsfonts}
\usepackage{mathrsfs}
\usepackage{tabularray}
\usepackage{physics}
\usepackage{epsfig,wrapfig}
\usepackage{subcaption}
\usepackage{graphicx}
\usepackage[bookmarks]{hyperref}
\usepackage[usenames,dvipsnames]{color}

\usepackage{units, overpic}
\usepackage[utf8]{inputenc}
\usepackage{rotating}
\usepackage{enumerate}
\usepackage{cancel}

\usepackage{soul}
\usepackage{xfrac}
\usepackage{multirow}

\usepackage{booktabs}
\usepackage{varwidth}
\usepackage{caption}
\newsavebox\tmpbox
\newcommand{\Natural}{\mathbb{N}}
\newcommand{\Real}{\mathbb{R}}

\newcommand{\re}{\mbox{Re}}

\newcommand{\rest}{\mathrm{rest}}
\newcommand{\inte}{\mathrm{int}}

\begin{document}

\title{Wave propagation through a spacetime containing thin concentric shells of matter}

\author{Rub\'en O. Acu\~na-C\'ardenas}

\affiliation{Instituto de F\'\i sica y Matem\'aticas,
Universidad Michoacana de San Nicol\'as de Hidalgo,
Edificio C-3, Ciudad Universitaria, 58040 Morelia, Michoac\'an, M\'exico.}

\author{Olivier Sarbach}

\affiliation{Instituto de F\'\i sica y Matem\'aticas,
Universidad Michoacana de San Nicol\'as de Hidalgo,
Edificio C-3, Ciudad Universitaria, 58040 Morelia, Michoac\'an, M\'exico.}
\affiliation{Departamento de Matem\'aticas Aplicadas y Sistemas, Universidad Aut\'onoma Metropolitana-Cuajimalpa, 05348, Cuajimalpa de Morelos, Ciudad de M\'exico, M\'exico.}

\author{Luca Tessieri}

\affiliation{Instituto de F\'\i sica y Matem\'aticas,
Universidad Michoacana de San Nicol\'as de Hidalgo,
Edificio C-3, Ciudad Universitaria, 58040 Morelia, Michoac\'an, M\'exico.}

\begin{abstract}
We investigate the transmission of scalar, electromagnetic, and linearized odd-parity gravitational waves in a static spacetime characterized by a spherical distribution of matter in the form of thin concentric equidistant shells of equal mass. These shells connect Schwarzschild spacetimes of different masses between themselves, and they satisfy the Israel junction conditions with a polytropic-type equation of state for the surface energy-momentum tensor. We assume that the central region has zero mass, and we verify that the resulting spacetime is stable with respect to small perturbations of the shell radii as long as the gravitational field is sufficiently weak.

We focus on the transmission of monochromatic waves emitted from the center and propagating through a succession of $N$ shells. To this purpose, we neglect the self-gravity of the waves and solve the Regge-Wheeler equation in the weak field limit of the background field. Analytical expressions for the transmission and reflection coefficients are obtained and their dependency on the frequency, the number of shells and their mutual distance is analyzed.
In particular, in the high-frequency limit, we observe that the reflection coefficient decays with the fourth power of the frequency. Increasing the number of shells initially produces oscillations in the transmission coefficient; however, as $N$ grows, this coefficient rapidly stabilizes at a constant positive value. We attribute this property to the fact that reflections are mainly determined by the surface density of the shells, which decreases as the inverse square of their radii.
\end{abstract}
\date{\today}

\maketitle
\tableofcontents

\section{Introduction}
\label{Sec:Intro}

The study of wave propagation in spacetime has intensified
in recent years, as a consequence of the groundbreaking
experimental observation of gravitational
waves~\cite{Art:abbott2016observation,Art:abbot2016Tests, Art:Abbott2017multimessenger} and important advances in numerical relativity simulations~\cite{Art:pretorius2009binary,Art:hinder2010current,Art:duez2018numerical}.
The detection of gravitational waves has provided new tools to
explore the universe~\cite{Art:Amaro2023Astrophysics}
and has increased the interest in the propagation of waves
of all kind in relativistic spacetimes and in their interaction
with different forms of matter, see for instance Refs.~\cite{Art:baym2017damping,Art:bishop2020effect,Art:modifications2021naidoo,Art:bishop2022effect,Art:Lieu2022damping,Art:Bishop2024Interaction}. However, in spite of contemporary advances,
our theoretical understanding of the influence that matter exerts
on wave propagation is still far from complete.

Initial steps in this direction have been performed by Esposito~\cite{Art:esposito1971interaction} and by Ehlers and collaborators~\cite{Art:Ehlers1987Propagation,Art:Ehlers1996GWsPF}, who demonstrated that an ideal fluid cannot extract energy from gravitational waves. However, when traversing a \emph{dissipative} fluid, gravitational waves are expected to be attenuated due to shear viscosity. This mechanism has been studied by several groups in the context of cosmology, see for instance Refs.~\cite{Art:hawking1966perturbations,Art:esposito1971absorption,Art:Madore1973theAbs, Art:Anile1978High, Art:prasanna1999propagation, Art:baym2017damping}. 
More recently, Bishop et al.~\cite{Art:bishop2020effect} analyzed the interaction of a gravitational wave produced by a source surrounded by a shell of dust matter. They found that the energy exchange is zero, in agreement with the conclusion from the aforementioned work, and that the wave undergoes modifications in its frequency, phase, and magnitude. In~\cite{Art:modifications2021naidoo} some astrophysical scenarios were proposed in which these effects could be important, including echoes in LIGO events or gravitational waves produced by core collapse supernovae. The last scenario was further explored in~\cite{Art:bishop2022effect} for a viscous fluid shell, and other potential applications such a gravitational wave heating of a shell of matter and the damping of primordial gravitational waves were discussed in Refs.~\cite{Art:kakkat2024gravitational} and \cite{Art:Bishop2024Interaction}.

In this article, we discuss an entirely different mechanism that could lead to the attenuation
of gravitational waves, namely the scattering of waves by thin shells of matter. 
These shells represent discontinuities in the spacetime
curvature and their simple structure makes it possible to simplify
the theoretical analysis of the interaction of matter with
gravitational, electromagnetic, and scalar waves.
Thin-shell models~\cite{Art:zloshchastiev1999barotropic,Art:khakshournia2002dynamics,Art:kijowski2006relativistic, Art:kijowski2010hamiltonian, Art:lemaitre2019equilibrium} have been applied to the analysis of various dynamical phenomena, including cosmological scenarios~\cite{Art:frolov1990black}, the stability of wormholes~\cite{Art:poisson1995thin,Art:Lobo2005Stability}, the study of the behavior of shells around black holes~\cite{Art:ASchell1990Frauendiener,Art:Brady1991Stability,Art:gonifmmode2002relshells, Art:Stability2014Pereira}, gravitational collapse~\cite{Art:gautreau1995gravitational, Art:adler2005simple}, and mass inflation~\cite{Art:poisson1990internal}.

More specifically, we study the propagation of scalar, electromagnetic,
and linearized odd-parity gravitational waves through a stationary
spacetime composed of thin spherical and concentrical 
shells of equal mass. These shells serve as interfaces connecting
Schwarzschild spacetimes of varying masses. We assume that the
Israel junction conditions~\cite{Art:israel1966singular, Art:israel1967gravitational,  Art:Mars1993Geometry} hold and that on each shell the surface energy-momentum tensor satisfies a polytropic equation of state.
The model under study has a massless central region, and its
stability against small perturbations of the
shell radii under weak gravitational fields is examined.
We focus on the transmission of monochromatic waves
originating from the center across an array of $N$ shells.
Applying the Regge-Wheeler (RW) equation to the case of a weak
gravitational background field and using the formalism introduced in Refs.~\cite{Art:buchman2006towards, Art:buchman2007improved},
we derive analytical expressions for the transmission and
reflection coefficients. These results allow us to determine
how the transmission properties depend on the wave frequency,
the number of thin shells, and the distance between neighboring
shells.
In particular, we show that the reflection
coefficient tends to vanish in the high-frequency limit and that
the transmission coefficient tends to a nonzero constant asymptotic value when
the number of shells increases.

Although in this article we do not consider even parity linearized gravitational waves which would interact with the dynamics of the shells, our results provide a detailed calculation for the attenuation of the outgoing odd-parity gravitational wave due to reflections at the thin shells. Furthermore, our findings also apply to (even and odd) electromagnetic waves and to scalar waves. The result for the second type of waves might be relevant for scalar field dark matter models which have attracted considerable interest in recent years (see, for instance, Refs.~\cite{Art:Marsh:2016darkmatter,Art:Ferreira2021darkmatter,book:jackson2023bosonicdarkmatter,Art:Matos2024darkmatter} for recent reviews).

The remainder of this work is organized as follows: in Sec.~\ref{Sec:Spacetime} we specify our background spacetime on which the wave propagation will be studied and discuss its stability and weak-field limit. In Sec.~\ref{Sec:RW} we construct approximate solutions of the RW equation for a spin $S$ field which are valid in the weak-field limit. Next, in Sec.~\ref{Sec:TransferMatrix} we work out the relevant matching conditions at each shell and introduce the transfer matrix formalism. The resulting reflection and transmission coefficients are computed in Sec.~\ref{Sec:TR} and their behavior is analyzed in detail in Sec.~\ref{Sec:Results}. Conclusions are drawn in Sec.~\ref{Sec:Conclusions}, where we also discuss possible extensions of this work. Finally, technical issues are considered in appendices~\ref{App:New_shells}, \ref{App:TranM_T_error}, \ref{App:EM_flux}, \ref{App:Aux_M}, and~\ref{App:WKB}.

In this work, we use geometrized units in which the gravitational constant $G$ and the speed of light $c$ are one, and we use the signature convention $(-,+,+,+)$ for the spacetime metric.

\section{Spacetime structure, stability, and weak-field limit}
\label{Sec:Spacetime}

\begin{figure}
\includegraphics[scale=0.35]{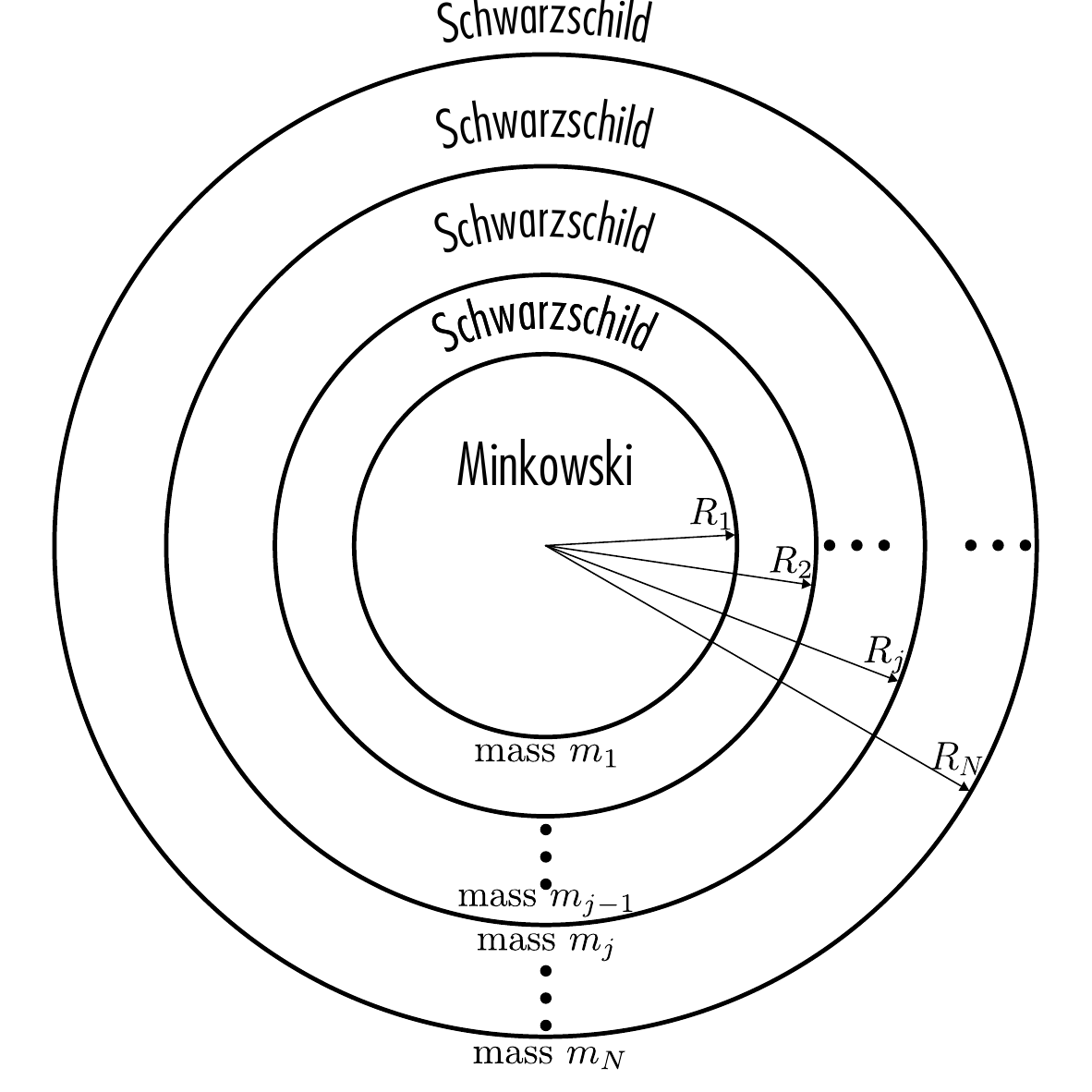}
\caption{Illustration for the spacetime structure consisting of $N$ thin concentric shells of matter.}
\label{Fig:spacetime-structure}
\end{figure}

In this section we specify the field equations for a spherically symmetric spacetime composed of a series of $N$ thin concentric shells with vacuum regions in between. According to Birkhoff's theorem, these shells must connect Schwarzschild spacetimes of distinct masses, and their dynamics must adhere to Israel's junction conditions~\cite{Art:israel1966singular,Art:israel1967gravitational, Art:Mars1993Geometry} with a suitable surface energy-momentum tensor which is assumed here to obey a polytropic-type equation of state. The central region is assumed to be flat (Minkowski) spacetime, see Fig.~\ref{Fig:spacetime-structure} for a depiction of the spacetime structure. In particular, we determine the equilibrium configurations leading to a static spacetime and analyze the stability of the resulting model. Finally, we take the weak-field limit of our model which greatly simplifies the subsequent analysis, and we show that the aforementioned stability conditions are automatically satisfied in this limit.

\subsection{Spacetime construction}

Our spacetime consists of $N+1$ copies of Schwarzschild spacetimes of increasing masses $0 = m_0  < m_1 < \ldots < m_N$. The spacetimes are glued together at three-dimensional hypersurfaces of the form $\Sigma_j := \{ r = \mathcal{R}_j(t) \}$ with the shells' radii satisfying the inequalities
\begin{equation}
0 < \mathcal{R}_1(t) < \mathcal{R}_2(t) < \ldots < \mathcal{R}_N(t) 
\end{equation}
for every time $t$. More precisely, the spacetime manifold is $\Real^4$, equipped with the metric
\begin{equation}
g = - \left( 1-\frac{2m(t,r)}{r} \right)dt^2 + \frac{dr^2}{1-\frac{2m(t,r)}{r}}
    + r^2 d\Omega^2 .
\label{Eq:metric_full_st}
\end{equation}
In Eq.~(\ref{Eq:metric_full_st}), $t\in\Real$ is the time, $r > 0$ the areal
radial coordinate, $d\Omega^2 = d\vartheta^2 + \sin^2{\vartheta} d\varphi^2$ is the standard metric on the unit two-sphere in spherical coordinates $(\vartheta,\varphi)$, and the mass function $m(t,r)$ is defined as follows
\begin{equation}
m(t,r) := m_j, \qquad \mathcal{R}_j(t) < r < \mathcal{R}_{j+1}(t),
\label{mass_fun}
\end{equation}
with $j=0,1,2,\ldots,N$. In Eq.~(\ref{mass_fun}) it is understood that
$\mathcal{R}_0 \equiv 0$ and $\mathcal{R}_{N+1} \equiv \infty$. Furthermore, we
assume that $2m(t,r) < r$ everywhere, i.e., that there are no horizons.

The dynamics of the shells is determined by Israel's junction conditions~\cite{Art:israel1966singular,Art:israel1967gravitational}, which ensures the satisfaction of Einstein's field equations at the interfaces $\Sigma_j$. In terms of the first and second fundamental forms $h_{ab}$ and $K_{ab}$ of $\Sigma_j$ these conditions are
\begin{align}
[h_{ab}] &= 0, \label{Eq:C_induced_metric}\\
h_{ab}[K]-[K_{ab}] &= 8\pi S_{ab} . \label{Eq:J_ext_curvature}
\end{align}
In Eqs.~(\ref{Eq:C_induced_metric}) and~(\ref{Eq:J_ext_curvature}) we use the
symbol $[X]$ to denote the discontinuity of any field $X$ across the surface $\Sigma_{j}$, i.e.,
\begin{equation}
[X] := X_{+}|_{\Sigma_j} - X_{-}|_{\Sigma_j} ,
\end{equation}
with $X_+$ and $X_-$ respectively representing the value of $X$ on the outer and inner border of $\Sigma_j$. The symbols
\begin{equation}
K := h^{ab} K_{ab}
\end{equation}
and $S_{ab}$ stand for the trace of $K_{ab}$ and for the surface energy-momentum tensor.
We will assume that $S_{ab}$ has the form associated to a perfect surface fluid:
\begin{equation}
S_{ab} = \left( \sigma + p \right)u_a u_b + p h_{ab},
\end{equation}
with $\sigma$, $p$, and $u^a$ respectively denoting the surface energy density,
the surface pressure, and the three-velocity of the fluid (taken to coincide with the
velocity of the comoving observers). As a consequence of the Codazzi-Mainardi equations and Einstein's equations, $S_{ab}$ is divergence-free.

The first and second fundamental form of the $j$-th shell~$\Sigma_j$, as embedded into the metric~(\ref{Eq:metric_full_st}) are given by
\begin{align}
\left. h_{ab} dx^a dx^b \right|_{\pm}  &= -d\tau^2 + \mathcal{R}^2 d\Omega^2,
\label{Eq:FirstFundamentalForm}\\
\left. K_{ab} dx^a dx^b \right|_{\pm} &= -\frac{1}{\mathcal{K}_\pm}\left( \ddot{\mathcal{R}} + \frac{m_\pm}{\mathcal{R}^2} \right) d\tau^2  + \mathcal{K}_\pm \mathcal{R} d\Omega^2,
\label{Eq:SecondFundamentalForm}
\end{align}
where we have set $\mathcal{K}_\pm := \sqrt{1 - 2m_\pm/\mathcal{R} + \dot{\mathcal{R}}^2}$ and suppressed the index $j$. Here and in the rest of this paper, the dot denotes derivation with respect to proper time~$\tau$ measured by observers which are comoving with the shells, rather than Schwarzschild time $t$.
Introducing the expressions~(\ref{Eq:FirstFundamentalForm},\ref{Eq:SecondFundamentalForm}) into Eqs.~(\ref{Eq:C_induced_metric},\ref{Eq:J_ext_curvature}) yields, after some manipulations \cite{Art:israel1966singular, Art:khorrami1991spherically, Art:gonifmmode2002relshells, Art:kijowski2006relativistic}
\begin{align}
\dot{\mathcal{R}}_j^{2} &= \frac{m_{\Sigma_j}^{2}}{4\mathcal{R}_j^{2}}
	+ \frac{m_{j} + m_{j-1}}{\mathcal{R}_j} 
	+\frac{\left(m_{j}-m_{j-1}\right)^{2}}{m_{\Sigma_j}^{2}} - 1, 
\label{Eq:Dot_R_j}\\
\dot{\sigma_j} &= -2\frac{\dot{\mathcal{R}_j}}{\mathcal{R}_j}(\sigma_j + p_j), 
\label{Eq:Dot_sigma_j}
\end{align}
where we have abbreviated
\begin{equation}
m_{\Sigma_j} := 4\pi\mathcal{R}_j^2\sigma_j.
\end{equation}
To close the system, one needs to specify an equation of state for the surface fluid. In this work, we propose a polytropic-type equation of state, for which
\begin{equation}
\sigma_j = \sigma_{\rest,j} + \sigma_{\inte,j},
\end{equation}
with vanishing pressure associated with the rest energy $p_{\rest,j} = 0$, and
a pressure associated with the internal energy of the form
\begin{equation}
p_j = \alpha \sigma_{\inte,j} .
\label{press}
\end{equation}
In the previous equation $\alpha$ is a constant which takes the same value for all the shells. In Sec.~\ref{eq_conf} and~\ref{stab} we will analyze the range of the physically acceptable values of $\alpha$. We require $\sigma_{\rest,j}$ and $\sigma_{\inte,j}$ to be positive. Note that the dominant (and hence also the weak) energy condition is satisfied if $|\alpha|\leq 1$ (see, for instance section 9.2 in Ref.~\cite{book:wald2010general}). It follows from Eq.~(\ref{Eq:Dot_sigma_j}) that the surface energy density evolves in time according to
\begin{equation}
\sigma_j(\tau) 
 = \sigma_{\rest,j}(0) \left( \frac{\mathcal{R}_j(0)}{\mathcal{R}_j(\tau)} \right)^2
 + \sigma_{\inte,j}(0) \left( \frac{\mathcal{R}_j(0)}{\mathcal{R}_j(\tau)}\right)^{2(1+\alpha)}.
\label{Eq:sigmaEvolution} 
\end{equation}
After substituting this expression into Eq.~(\ref{Eq:Dot_R_j}) one obtains the equation of motion for the $j$-th shell which has the same form as the equation for a one-dimensional mechanical particle in an external potential. In what follows we explicitly derive the form of this dynamical equation and determine the equilibrium points and their stability.

\subsection{Equilibrium configurations}
\label{eq_conf}

In this subsection we determine the conditions under which the shells are in equilibrium (so that the resulting spacetime is static) and, furthermore, they are stable with respect to small radial perturbations. We shall denote the radii of the static shells with the italic symbols $R_j$ and introduce the new variables $z_j := \mathcal{R}_j/R_j$. Using Eqs.~(\ref{Eq:Dot_R_j},\ref{Eq:sigmaEvolution}) we can rewrite the $j$-th shell's equation of motion as
\begin{equation}
R_j^2 \dot{z}_j + V_j(z_j) = 0, 
\end{equation}
with the effective potential
\begin{widetext}
\begin{equation}
V_j(z) := 1 
- \left[ 1 - \frac{(k_j^+){^2} + (k_j^-)^2}{2} \right]\frac{1}{z} 
- \frac{ \left(a_{\rest,j} + a_{\inte,j} z^{-2\alpha}\right)^{2} }{z^2}
- \frac{\left[ (k_j^-)^2 - (k_j^+)^2\right]^{2}}{16 
\left( a_{\rest,j} + a_{\inte,j}z^{-2\alpha}\right)^{2}} .
\label{poten}
\end{equation}
\end{widetext}
To simplify the notation, in Eq.~(\ref{poten}) we have introduced the dimensionless
variables $a_{\rest,j} := 2\pi R_j\sigma_{\rest,j}$, $a_{\inte,j} := 2\pi R_j\sigma_{\inte,j}$, and
\begin{equation}
k_j^{-} := \sqrt{1 - \frac{2m_{j-1}}{R_j}}, \\ \quad
k_j^{+} := \sqrt{1-\frac{2m_{j}}{R_j}}.
\end{equation}

The equilibrium points are determined by the conditions $V_j(1) = V_j'(1) = 0$. After performing some tedious but straightforward calculations, one
obtains
\begin{align}
k_j^{-} - k_j^+
&= 4\pi R_j(\sigma_{\rest,j} + \sigma_{\inte,j}),	
\label{Eq:equilibrium_one}\\
k_j^{-} k_j^{+}
&= \frac{\sigma_{\rest,j} + \sigma_{\inte,j}}
{\sigma_{\rest,j} + \sigma_{\inte,j}(1+4\alpha)}.
\label{Eq:equilibrium_two}
\end{align}
Since $k_j^\pm < 1$ this requires, in particular, that the constant $\alpha$ in Eq.~(\ref{press}) be positive, $\alpha > 0$. The physical content of equilibrium conditions~(\ref{Eq:equilibrium_one}) and~(\ref{Eq:equilibrium_two}) will be discussed in Sec.~\ref{SubSec:WeakField}, where the weak-field
limit is considered.

\subsection{Stability}
\label{stab}

\begin{figure}[htp]

\subfloat{%
  \includegraphics[clip,width=\columnwidth]{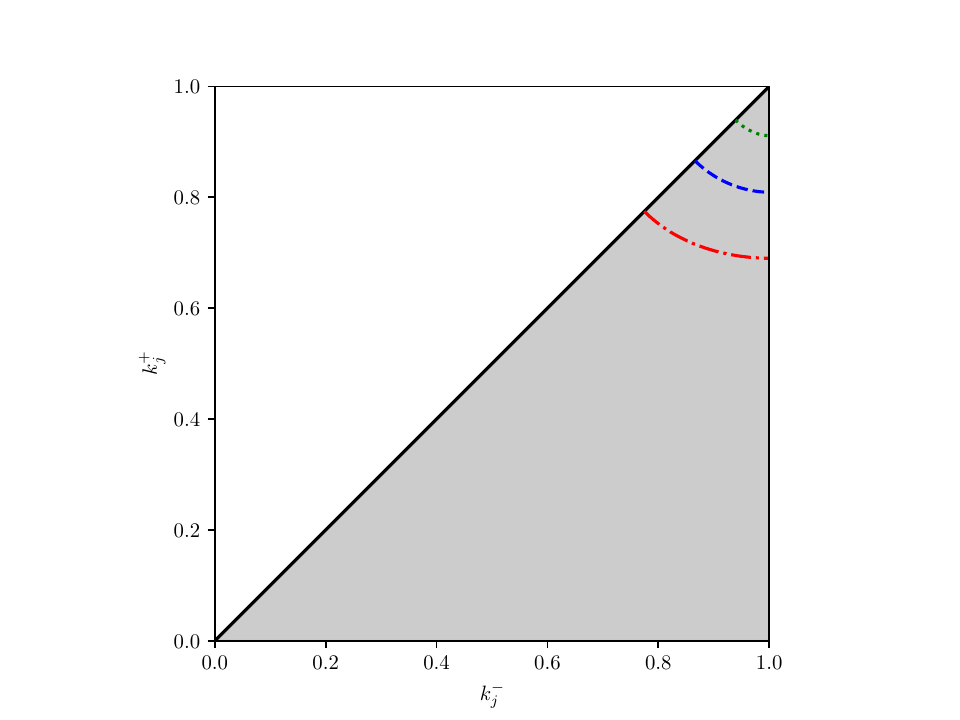}%
}

\subfloat{%
  \includegraphics[clip,width=\columnwidth]{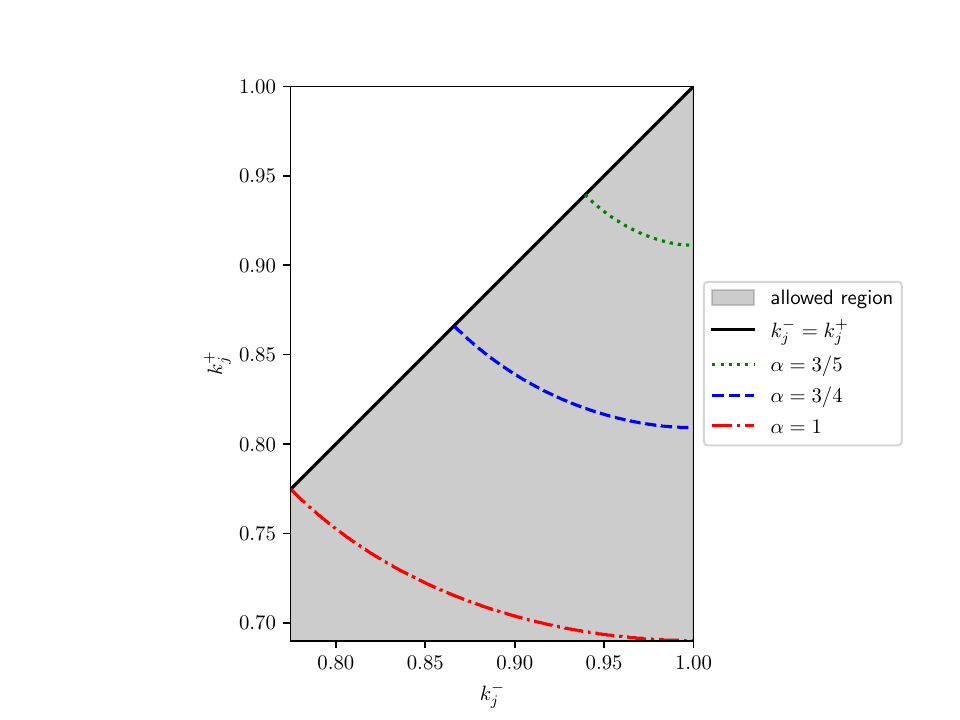}%
}
\caption{Stability regions within the triangular zone $0 < k_j^{-} < k_j^{+}  < 1$ (in gray) associated with physically allowed realizations for a static shell in the $(k_j^{-},k_j^{+})$ plane. The solid black line $k_j^{-} = k_j^{+}$ corresponds to the limit of shells with vanishing surface energy density. Stable shells lie in the region above  the colored dotted, dashed, or dash-dotted curves, indicated here for values of $\alpha$ equal to $3/5$, $3/4$ and $1$, respectively. The bottom panel shows an enlargement of the top panel, to make the stability regions more visible.}
\label{Fig:stability_reg}
\end{figure}

The stability of the static configurations with respect to radial fluctuations of the shells can be determined by analyzing the sign of the second derivative of the effective potential $V_j$ at the equilibrium point. More precisely, the condition $V_j''(1) > 0$ guarantees stability. In our analysis of the stability of spherical shells connecting two Schwarzschild spacetimes we follow the previous works on the subject~\cite{Art:Brady1991Stability,Art:kijowski2006relativistic,Art:lemaitre2019equilibrium}. Computing the second derivative of $V_j$ and using the equilibrium conditions~(\ref{Eq:equilibrium_one},\ref{Eq:equilibrium_two}) to eliminate $a_{\rest,j}$ and $a_{\inte,j}$ leads to
\begin{align}
& V_j''(1) \nonumber\\
 &= 2 + 2\alpha - (1+4\alpha)\frac{k_j^{+} k_j^{-}}{2}
 - \frac{(k_j^{+})^2 + (k_j^{-})^2 + k_j^{+} k_j^{-}}{2(k_j^{+} k_j^{-})^2}.
\label{Eq:2thDer_efective_Pot}
\end{align}
This expression allows one to determine the stability region in terms of the two parameters $k_j^{\pm}$ which satisfy $0 < k_j^+ < k_j^- < 1$. In the weak-field regime, $k_j^- = 1 - m_{j-1}/R_j + {\cal O}(\varepsilon^2)$ and $k_j^+ = 1 - m_j/R_j + {\cal O}(\varepsilon^2)$ with $\varepsilon = m_j/R_j$, and a short calculation gives
\begin{equation}
V_j''(1) = (2\alpha-1)\frac{m_{j-1}+m_j}{R_j} + {\cal O}(\varepsilon^2),
\end{equation}
which is positive as long as $\alpha > 1/2$. In the limit $k_j^- = k_j^+$, which corresponds to shells with vanishing surface energy density, one finds that the second derivative is positive if and only if $\alpha > 1/2$ and
\begin{equation}
\sqrt{\frac{3}{1+4\alpha}} < k_j^+ < 1.
\end{equation}
Finally, in the limit $k_j^- = 1$ one finds that $V_j''(1) > 0$ if and only if $\alpha > 1/2$ and
\begin{equation}
\frac{1 + \sqrt{2(1+2\alpha)}}{1+4\alpha} < k_j^+ < 1.
\end{equation}
Figure~\ref{Fig:stability_reg} shows the stability regions for different values of $\alpha$.

\subsection{Weak-field limit and specific model}
\label{SubSec:WeakField}
 
To avoid mathematical complications which could obscure the physical results, in what follows we shall restrict ourselves to spacetimes with {\em weak} gravitational field and whose shells have surface energy dominated by their rest component, i.e.,
\begin{eqnarray}
\frac{2m_j}{R_j}\ll 1,
\label{Eq:WeakFieldLimit}\\
\sigma_{\inte,j} \ll \sigma_{\rest,j}.
\label{Eq:LowTemperatureLimit}
\end{eqnarray}
As shown in the previous subsection, the conditions~(\ref{Eq:WeakFieldLimit},\ref{Eq:LowTemperatureLimit}) ensures that the configuration is stable as long as $\alpha > 1/2$. In this limit, Eqs.~\eqref{Eq:equilibrium_one} and \eqref{Eq:equilibrium_two} reduce to the much simpler conditions
\begin{align}
m_{j}-m_{j-1} & = 4\pi R_j^2 \sigma_{\rest,j},	
\label{Eq:shell_mass}\\
m_{j}+m_{j-1} & = 4R_{j}\frac{p_{j}}{\sigma_{\rest,j}}.
\label{Eq:hydrostatic_condition}
\end{align}
The first condition has an obvious interpretation; it states that the masses $m_{j}$ and $m_{j-1}$ differ by the total mass of the $j$-th shell. The second condition originates from the requirement of hydrostatic equilibrium, and it can be derived using purely Newtonian arguments (see
Refs.~\cite{Art:lemaitre2019equilibrium,Art:gonifmmode2002relshells} and Appendix~\ref{App:New_shells} for a more detailed discussion).

From now on, we shall focus on a system of thin equidistant shells of equal masses. We would like to stress that our techniques make possible to analyze models of a more general nature; however, we restrict our attention to this specific case for the sake of simplicity. Hence, we assume that the masses $m_j$ and radii $R_j$ are of the form
\begin{equation}
m_j = j\Delta m,\qquad
R_j = R_1 + (j-1)\Delta R,
\label{Eq:mj_rj}
\end{equation}
for $j=1,2,\ldots,N$. The positive constants $\Delta m$, $\Delta R$, and $R_1$ in Eq.~(\ref{Eq:mj_rj}) are required to satisfy the conditions
\begin{equation}
\frac{\Delta m}{R_1} \ll 1,\qquad
\frac{\Delta m}{\Delta R} \ll 1,
\label{Eq:WeakFieldConditions}
\end{equation}
for consistency with the weak-field limit~\eqref{Eq:WeakFieldLimit}. It follows from
Eqs.~(\ref{Eq:shell_mass}) and~(\ref{Eq:hydrostatic_condition}) that
\begin{equation}
\sigma_{\rest,j} = \frac{\Delta m}{4\pi R_j^2},\qquad
p_j = \frac{(2j-1)(\Delta m)^2}{16\pi R_j^3},
\label{Eq:rhoj_pj}
\end{equation}
for $j=1,2,\ldots,N$. For large $j$ the ratio $p_j/\sigma_{\rest,j}$, which is proportional to the temperature of the shell, and the compactness ratio $2m_j/R_j$ are given by
\begin{align}
\lim\limits_{j\to \infty}\frac{p_j}{\sigma_{\rest,j}} &= \frac{\Delta m}{2 \Delta R} \ll 1,\\
\lim\limits_{j\to \infty}\frac{2m_j}{R_j} &= 2\frac{\Delta m}{\Delta R} \ll 1.
\end{align}
The first condition implies that we are working in the low temperature limit, whereas the second one allows us to consider spacetimes with an arbitrarily high number of shells within the weak-field approximation. For large $N$, these shells have compactness ratio and temperature approaching a constant value as $j$ becomes large. In contrast to this, note that the surface mass density decays as $1/j^2$ for large $j$. This property will turn out to be important when we analyze the transmission of waves through a large number of shells (see Sec.~\ref{Sec:Results}).

To sum up, Eq.~\eqref{Eq:mj_rj} defines a static spacetime consisting of $N$ spherical concentric and equidistant shells of equal mass. Provided conditions~\eqref{Eq:WeakFieldConditions} are satisfied, this spacetime is stable with respect to small fluctuations of the shells' radii and the gravitational field is weak everywhere. In the rest of this article we shall study the propagation of linear waves emanating from the central region.

\section{Approximate solutions of the Regge-Wheeler equation}
\label{Sec:RW}

The propagation of scalar, electromagnetic and linearized gravitational waves in a spherically symmetric spacetime background is described by the RW equation~\cite{Art:Stability1957RW} and its generalizations (see, for instance, Refs.~\cite{Art:Mathematical1984Chandrasekhar,Art:linearpert2013Chaverra} and references therein). The RW equation is a wave-type ``master" equation for a gauge-invariant field $\Phi$ from which the original (scalar, electromagnetic or linearized metric) field can be reconstructed. In this section we construct explicit monochromatic incoming and outgoing solutions of the RW master equation assuming a background spacetime of the form considered in the previous section and depicted in Fig.~\ref{Fig:spacetime-structure}. Since the RW equation belongs to the confluent Heun class (see, for instance~\cite{Art:Exact2006Fiziev}), it is in principle possible to construct \emph{exact} solutions which, however, present a significant interpretative challenge when one considers the nature of incoming and outgoing waves at infinity. For this reason, we adopt a different approach in this article: we focus on the weak-field limit and perform an expansion of the incoming and outgoing solutions in terms of $2m/r$. For a Schwarzschild spacetime, such an expansion can be worked out in a systematic way and leads to a convergent series when done appropriately~\cite{Art:bardeen1973radiation}. However, for the purpose of this article it is sufficient to consider only the zeroth and the first-order terms in $2m/r$, since higher-order corrections have already been neglected when taking the weak-field limit in Subsec.~\ref{SubSec:WeakField}.

When decomposed into spherical harmonics, the RW master equation for a field of spin $S$ propagating on the spacetime background~(\ref{Eq:metric_full_st}) reads
\begin{widetext}
\begin{equation}
\left[\frac{\partial^{2}}{\partial t^{2}}
- \frac{\partial^{2}}{\partial r_{*}^{2}}
+ \left( 1 - \frac{2m(r)}{r} \right)
\left( \frac{\ell(\ell+1)}{r^2}
+ (1-S^{2})\frac{2m(r)}{r^3} 
\right)\right]\Phi_\ell(t,r) = 0 .
\label{Eq:ReggeWheeler}
\end{equation}
\end{widetext}
In Eq.~(\ref{Eq:ReggeWheeler}), $\ell\geq S$ denotes the total angular momentum of the field and we do not label the field $\Phi_\ell$ with the magnetic quantum number $m$ since it does not enter the equation. The number $S$ takes the values
\begin{equation}
S := 
\begin{cases}
0 & \textrm{scalar},\\
1 & \textrm{electromagnetic,}\\
2 & \textrm{odd-parity linearized gravitational},
\end{cases}
\end{equation}
and $r_*$ is the tortoise coordinate, defined as
\begin{equation}
r_* := \int_{r_0}^r \left(1-\frac{2m(x)}{x} \right)^{-1} dx,
\label{Eq:tortoise_coordinate}
\end{equation}
with an arbitrary constant $r_0 > 0$. Note that in Eqs.~(\ref{Eq:ReggeWheeler}) and (\ref{Eq:tortoise_coordinate}) we have written $m(r)$ rather than $m(t,r)$ since the spacetime is static. We will stick to this convention in the rest of the paper. Since the mass function vanishes in the central region we can take $r_0 = 0$; evaluation of the resulting integral yields
\begin{multline}
r_* = 
r +
2\sum_{k=1}^{j-1}m_{k}\log{\left(
\frac{R_{k+1}-2m_k}{R_k-2m_k}
\right)} \\	+
2m_{j}\log{\left(
\frac{r - 2m_{j}}{R_{j}-2m_{j}}
\right)}	,
\label{Eq:explicit_tortoise_coo}
\end{multline}
for $R_j < r < R_{j+1}$. In Eq.~(\ref{Eq:explicit_tortoise_coo}) and in the rest of this paper, we adopt the convention that a sum is zero if the upper bound is smaller than the lower one. Hence, $r_* = r$ inside the first shell ($0 < r < R_1$) and $r_* = r + 2m_1 \log\left(\frac{r - 2m_1}{R_1 - 2m_1}\right)$ in the region $R_1 < r < R_2$ between the first and second shell. Note that $r_*$ is a continuous function of $r$; however, its first derivative jumps at each shell.

In what follows, we represent the solution to Eq.~\eqref{Eq:ReggeWheeler} as a sum of incoming ($\Phi_{\nwarrow,\ell}$) and outgoing ($\Phi_{\nearrow,\ell}$) wave functions, with respective amplitudes $\Lambda_\ell$ and $\Upsilon_\ell$:
\begin{equation}
\Phi_\ell (t,r) = \Lambda_{\ell} \Phi_{\nwarrow,\ell}(t,r)
+\Upsilon_{\ell} \Phi_{\nearrow,\ell}(t,r).
\label{Eq:Complete_WaveF}
\end{equation}
In the next four subsections we construct these functions in the region between two shells, where $m$ is constant, whereas the matching conditions for the function $\Phi_\ell$ across the shells are worked out in Subsec.~\ref{SubSec:WaveFunctionMatching}.

\subsection{Perturbative setting}
\label{SubSec:Setting}

As mentioned above, we construct the functions $\Phi_{\nwarrow,\ell}$ and $\Phi_{\nearrow,\ell}$ using a perturbative approach based on the method developed in~\cite{Art:bardeen1973radiation}. We follow closely the presentation and notation introduced in~\cite{Art:buchman2006towards,Art:buchman2007improved}.  However, whereas the last two references only consider the cases $S=\ell=2$ and $S=\ell=0$, here we generalize these results to arbitrary values of $S$ and $\ell$.

As a first step, we change the Schwarzschild coordinates $(t,r)$ to the new coordinates $(\tau,\rho)$, defined by\footnote{Note that these coordinates are related to the outgoing Eddington-Finkelstein coordinates $(u,r) = (t-r_*,r)$.}
\begin{equation}
\tau := t - r_* + r, \qquad
\rho := r.
\end{equation}
This allows one to rewrite the RW master equation~\eqref{Eq:ReggeWheeler} in the form
\begin{equation}
\mathcal{L}_\ell\Phi_{\ell}(\tau,\rho) 
 = -\frac{2m}{\rho}\mathcal{B}\Phi_{\ell}(\tau,\rho),
\label{Eq:ReggeWheeler_tau_rho}
\end{equation}
where $\mathcal{L}_\ell$ denotes the (spherically reduced) wave operator in flat spacetime,
\begin{equation}
\mathcal{L}_\ell := \frac{\partial^2}{\partial\tau^2} - \frac{\partial^2}{\partial\rho^2}
 +\frac{\ell(\ell+1)}{\rho^{2}}, 
\label{Eq:L_operator}
\end{equation}
and the operator $\mathcal{B}$ is defined as
\begin{equation}
\mathcal{B} := \left(\frac{\partial}{\partial\tau} + \frac{\partial}{\partial\rho} \right)^2
 -\frac{1}{\rho}\left(\frac{\partial}{\partial\tau} + \frac{\partial}{\partial\rho} \right)
 + \frac{1-S^2}{\rho^2}.
\label{Eq:B_operator}
\end{equation}
The representation~(\ref{Eq:ReggeWheeler_tau_rho}) in which the small parameter $2m/\rho$ appears on the right-hand side of the equation allows one to seek the outgoing solution in the form
\begin{equation}
\Phi_{\nearrow,\ell}(\tau,\rho) = h_{\nearrow,\ell}(\tau,\rho) 
 + \sum_{k=1}^{\infty}\left(\frac{2m}{R}\right)^{k}g_{k,\ell}(\tau,\rho) .
\label{Eq:perturbative_ansatz}
\end{equation}
In Eq.~(\ref{Eq:perturbative_ansatz}) $R$ is a typical radius ($R\approx R_j$ for our problem), the functions $g_{k,\ell}$ are to be determined, and $h_{\nearrow,\ell}$ is the outgoing solution of the flat wave equation which is constructed in the next subsection. Substituting the ansatz~\eqref{Eq:perturbative_ansatz} into Eq.~\eqref{Eq:ReggeWheeler_tau_rho} results in an infinite hierarchy of wave equations
\begin{equation}
\mathcal{L}_\ell g_{k,\ell} = -\frac{R}{\rho}\mathcal{B} g_{k-1,\ell}, \qquad
k = 1,2,3,\ldots,
\label{Eq:Hierarchy_Dequations}
\end{equation}
with $g_0 := h_{\nearrow,\ell}$. As stated previously, in this work we only solve the first of these equations.

\subsection{Flat spacetime outgoing solution}
\label{SubSec:ZerothOrder}

The flat spacetime outgoing solution can be generated from a smooth function $U_0$ depending only on (minus) retarded time $-u = r_* - t = \rho - \tau$. For $\ell=0$ it is sufficient to define $h_{\nearrow,\ell}(\tau,\rho) := U_0(\rho-\tau)$, since $\mathcal{L}_0 U_0(\rho-\tau) = 0$. For $\ell > 0$ the solutions can be obtained by successive application of the ``creation'' operators $a_\ell^\dagger$, where
\begin{equation}
a_\ell := \frac{\partial}{\partial\rho} + \frac{\ell}{\rho},\qquad
a_\ell^\dagger := -\frac{\partial}{\partial\rho} + \frac{\ell}{\rho}
\end{equation}
satisfy
\begin{equation}
a_{\ell+1} a_{\ell+1}^\dagger = a_\ell^\dagger a_\ell 
 = -\frac{\partial^2}{\partial\rho^2} + \frac{\ell(\ell+1)}{\rho^2},
\end{equation}
from which it follows that $\mathcal{L}_\ell a_\ell^\dagger a_{\ell-1}^\dagger \cdots a_1^\dagger = a_\ell^\dagger a_{\ell-1}^\dagger \cdots a_1^\dagger\mathcal{L}_0$. Therefore, the function
\begin{align}
h_{\nearrow,\ell}(\tau,\rho)
 &:= a_\ell^\dagger a_{\ell-1}^\dagger \cdots a_1^\dagger U_0(\rho-\tau) \nonumber\\
 & = \sum_{k=0}^{\ell}\frac{(\ell+k)!}{k!(\ell-k)!}\frac{(-1)^{\ell}}{(-2\rho)^{k}}\frac{d^{\ell-k}}{d\rho^{\ell-k}}U_{0}(\rho-\tau)
\label{Eq:homogeneous_waveF}
\end{align}
satisfies the flat wave equation $\mathcal{L}_\ell h_{\nearrow,\ell} = 0$.

Similarly, incoming solutions can be constructed by changing the sign of $\tau$. For the next subsection the following solution will play an important role:
\begin{align}
& K_\ell(\tau,\rho,x) := a_\ell^\dagger a_{\ell-1}^\dagger \cdots a_1^\dagger \frac{1}{(\tau + \rho + x)^2}\nonumber\\
 &= \frac{1}{(2\rho)^{\ell+2}}\sum_{k=0}^{\ell}\frac{(2\ell-k)!(k+1)}{(\ell-k)!}\left(\frac{2\rho}{\tau + \rho + x}\right)^{k+2}.
\label{Eq:Kell}
\end{align}
Using the binomial identity
\begin{equation}
\sum\limits_{k=0}^p \binom{\ell+k}{k} = \binom{\ell+1 + p}{p},\qquad
p = 0,1,2,\ldots,
\end{equation}
it is not difficult to verify that
\begin{equation}
\left. 
\left(\frac{\partial}{\partial\tau} + \frac{\partial}{\partial\rho} \right)K_\ell(\tau,\rho,x)
\right|_{x = \rho-\tau} = -\frac{(2\ell+1)!}{2^{\ell+1}\ell!}\frac{1}{\rho^{\ell+3}},
\end{equation}
from which one obtains the following identity:
\begin{equation}
\mathcal{L}_\ell\int\limits_{\rho-\tau}^\infty K_\ell(\tau,\rho,x) U_0(x) dx = -\frac{(2\ell+1)!}{2^\ell\ell!}\frac{U_0(\rho-\tau)}{\rho^{\ell+3}},
\end{equation}
valid for any smooth function $U_0(x)$ which is bounded for $x\to \infty$. This identity will be useful in the next subsection, when computing the first-order correction terms.

\subsection{Outgoing solution including first-order correction terms}
\label{SubSec:FirstOrder}

The first-oder correction term $g_{1,\ell}$ is determined by Eq.~(\ref{Eq:Hierarchy_Dequations}) for $k=1$, that is,
\begin{equation}
\mathcal{L}_\ell g_{1,\ell} = -\frac{R}{\rho}\mathcal{B} h_{\nearrow,\ell},
\label{Eq:FirstOrderEq}
\end{equation}
where $h_{\nearrow,\ell}$ is given by Eq.~(\ref{Eq:homogeneous_waveF}). Using Eqs.~(\ref{Eq:B_operator}) and (\ref{Eq:homogeneous_waveF}) one finds
\begin{align}
& \mathcal{B} h_{\nearrow,\ell}(\tau,\rho) \nonumber\\
 &= \frac{(-1)^\ell}{\rho^2}\sum\limits_{k=0}^\ell \frac{(\ell+k)!}{k!(\ell-k)!}
 \frac{k(k+2) + 1 - S^2}{(-2\rho)^k}\frac{d^{\ell-k}}{d\rho^{\ell-k}}U_{0}(\rho-\tau).
\end{align}

Motivated by this form and the results in section~5.2 of Ref.~\cite{Art:buchman2006towards} and the appendix in Ref.~\cite{Art:buchman2007improved}, we propose the following ansatz for $g_{1,\ell}(\tau,\rho)$:
\begin{equation}
g_{1,\ell}(\tau,\rho) = R\left(\phi_{\ell}(\tau,\rho) + \beta\int_{\rho-\tau}^{\infty}K_{\ell}\left(\tau,\rho,x\right)U_{0}(x)dx\right),
\label{Eq:Ansatz_g1}
\end{equation}
where
\begin{equation}
\phi_\ell(\tau,\rho)
  := \sum_{k=0}^{\ell}\frac{\gamma_{\ell k}}{(k+1)!}\frac{(-1)^{\ell+1}}{(-2\rho)^{k+1}}
\frac{d^{\ell-k}}{d\rho^{\ell-k}} U_{0}(\rho-\tau),
\end{equation}
the integral kernel $K_\ell$ is given by Eq.~(\ref{Eq:Kell}), and the constants $\beta$ and $\gamma_{\ell k}$ need to be adjusted. Substituting this ansatz into Eq.~\eqref{Eq:FirstOrderEq} yields $\beta=1$ and $\gamma_{\ell,0} = 0$, whereas the remaining constants are obtained from the recurrence relation
\begin{equation}
\gamma_{\ell k} =
\frac{2k(\ell+k-1)!}{(\ell-k+1)!}(k^2-S^2) + (\ell-k)(\ell+k+1)\gamma_{\ell,k-1},
\end{equation}
for $k=1,2,\ldots,\ell$. Using the previous results, one can write the outgoing wave solution as
\begin{widetext}
\begin{multline}
\Phi_{\nearrow,\ell}(\tau,\rho) = 
(-1)^\ell\sum_{k=0}^\ell\left\{ \left[ \frac{(\ell+k)!}{k!(\ell-k)!} 
 + \frac{m}{\rho}\frac{\gamma_{\ell k}}{(k+1)!} \right]
 \frac{1}{(-2\rho)^k}\frac{d^{\ell-k}}{d\rho^{\ell-k}}U_0(\rho-\tau)\right. \\
\left. + \frac{m}{\rho}\frac{(2\ell-k)!(k+1)}{(\ell-k)! (-2\rho)^\ell}
\int_1^\infty U_0(2\rho y-\tau-\rho)\frac{dy}{y^{k+2}} \right\} 
 + {\cal O}\left( \frac{2m}{R} \right)^2.
\label{Eq:Outgoing}
\end{multline}
\end{widetext}
The terms in the first line include the first-order correction in $2m/R$ from the curvature of the background and obey Huygens' principle, whereas the terms on the second line describe the leading-order effects from the backscatter. Notice that these terms are present for all $\ell\geq S$; hence backscatter is always present for linear waves of spin $S$ fields on a Schwarzschild background. For dipolar electromagnetic radiation, $\ell=S=1$ and one finds $\gamma_{1,0} = \gamma_{1,1} = 0$, whereas for quadrupolar linearized gravitational waves one has $\ell=S=2$, and therefore
\begin{equation}
\gamma_{2,0} = \gamma_{2,2} = 0,\qquad
\gamma_{2,1} = -6,
\end{equation}
and Eq.~(\ref{Eq:Outgoing}) coincides with the outgoing solution on page~6735 of Ref.~\cite{Art:buchman2006towards}.

In what follows, we consider a monochromatic plane wave of frequency $\omega$, for which the function $U_0$ is given by
\begin{equation}
U_0(x) := \frac{e^{-sx}}{s^\ell},\qquad s:= -i\omega,
\end{equation}
such that
\begin{equation}
U_0(\rho - \tau) = \frac{1}{s^\ell} e^{-i\omega(\tau - \rho)}
 = \frac{1}{s^\ell} e^{-i\omega(t - r_*)}.
\end{equation}
Substituting this into Eq.~\eqref{Eq:Outgoing} and expressing the result in terms of the original coordinates $(t,r)$ yields the outgoing solution
\begin{equation}
\Phi_{\nearrow,\ell}\left(t,r\right) = e^{-i\omega t}X_{\nearrow,\ell}(r),
\label{Eq:Phi_ne}
\end{equation}
with
\begin{widetext}
\begin{equation}
X_{\nearrow,\ell}(r) = e^{i\omega r_*}
\sum_{k=0}^{\ell}\left\{ \left[ \frac{(\ell+k)!}{k!(\ell-k)!} 
 + \frac{m}{r}\frac{\gamma_{\ell k}}{(k+1)!} \right]\frac{1}{(2sr)^k}
 + \frac{m}{r}\frac{(2\ell-k)!(k+1)}{(\ell-k)!(2sr)^{\ell}} e^{2sr}E_{k+2}(2sr) \right\}
 + {\cal O}\left( \frac{2m}{r} \right)^2,
\label{Eq:RadialW_sol}
\end{equation}
\end{widetext}
where we recall that $s = -i\omega$ and where $E_p$ refers to the generalized exponential integrals, defined by~\cite[pg.~185]{book:olver2010nist}
\begin{equation}
E_p(z) := \int\limits_1^\infty \frac{e^{-z y}}{y^p} dy,\qquad
p = 0,1,2,\ldots
\end{equation}
They are holomorphic for $\re(z) > 0$, finite for $\re(z)\geq 0$ when $p\geq 2$, satisfy the recurrence relation
\begin{equation}
p E_{p+1}(z) = e^{-z} - z E_p(z),
\end{equation}
and behave as $E_p(z) \sim e^{-z}/z$ for $z\to \infty$.

This concludes our discussion of the outgoing solution including the first-order correction terms in $2m/r$.

\subsection{Incoming solution including first-order correction terms}
\label{SubSec:FirstOrderIn}

Taking advantage of the fact that the RW equation~\eqref{Eq:ReggeWheeler} is invariant with respect to the inversion of time $t\mapsto -t$, the incoming solution $\Phi_{\nwarrow,\ell}$ can be obtained from the outgoing one as follows:
\begin{equation}
\Phi_{\nwarrow,\ell}(t,r) = \overline{\Phi_{\nearrow,\ell}(-t,r)} 
 = e^{-i\omega t} \overline{X_{\nearrow,\ell}(r)}.
 \label{cc}
\end{equation}
Therefore, the incoming solution is obtained by replacing $X_{\nearrow,\ell}$ by its complex conjugate. Note that in Eq.~(\ref{cc}) we use the symbol $\overline{(\cdots)}$ to denote the complex
conjugate. We shall follow this convention throughout this paper.

Accordingly, Eq.~\eqref{Eq:Complete_WaveF} can be rewritten in the form
\begin{equation}
\Phi_\ell (t,r) = e^{-i\omega t}X_\ell(r),\quad
X_\ell(r) := \Lambda_\ell X_{\nwarrow,\ell}(r) 
 + \Upsilon_\ell X_{\nearrow,\ell}(r)
\label{Eq:MonochromaticInOut}
\end{equation}
with $X_{\nearrow,\ell}$ given by Eq.~\eqref{Eq:RadialW_sol} and
\begin{equation}
X_{\nwarrow,\ell}(r) = \overline{X_{\nearrow,\ell}(r)}.
\label{Eq:XnenwSym}
\end{equation}

\subsection{Matching conditions at the shells}
\label{SubSec:WaveFunctionMatching}

So far, the incoming and outgoing solutions have been constructed in the regions between the shells, where $m$ is constant. Here, we discuss the matching conditions that allow one to join the solutions at the shells. We shall use the symbol $X_{\ell,j}(r)$ for the solution~\eqref{Eq:MonochromaticInOut} within the region $R_j < r < R_{j+1}$ and we will represent the amplitudes of the corresponding in- and out-going waves with $\Lambda_{\ell,j}$ and $\Upsilon_{\ell,j}$. Note that the total mass inside this region is $m = m_j$. The matching conditions are understood best when looking at the RW equation~\eqref{Eq:ReggeWheeler} in terms of the coordinates $(t,r_*)$. Recall that $r_*$ is a continuous monotonously increasing function of $r$, such that $r$ also depends continuously on $r_*$. However, the function $m(r)$ jumps at each shell; thus, when introducing the monochromatic ansatz $\Phi_\ell (t,r) = e^{-i\omega t}X_\ell(r)$, one obtains a time-independent Schr\"odinger equation with a potential that jumps at each shell. As is well-known, for such problems the correct matching conditions are the requirements that both $X_\ell(r)$ and its first derivative (with respect to $r_*$) must be continuous, which leads to
\begin{equation}
X_{\ell,j}(R_j) = X_{\ell,j-1}(R_j),
\label{Eq:Matching1}
\end{equation}
and
\begin{equation}
\left.\frac{dX_{\ell,j}(r)}{dr_*}\right|_{r=R_j} 
= \left.\frac{dX_{\ell,j-1}(r)}{dr_*}\right|_{r=R_j}
\label{Eq:Matching2}
\end{equation}
for all $j=1,2,\ldots,N$. However, note that the derivative of $X_\ell(r)$ with respect to the areal radius $r$ jumps since $dr_*/dr$ involves the function $m(r)$ (see Eq.~\eqref{Eq:tortoise_coordinate}).

Using Eq.~\eqref{Eq:MonochromaticInOut}, the matching conditions~\eqref{Eq:Matching1},\eqref{Eq:Matching2} can be reformulated in terms of the coefficients $\Lambda_{\ell,j}$ and $\Upsilon_{\ell,j}$ as follows:
\begin{equation}
\mathbb{D}_{\ell,j}(R_{j})\left(\begin{array}{c}
\Upsilon_{\ell,j}\\
\Lambda_{\ell,j}
\end{array}\right) = 
\mathbb{D}_{\ell,j-1}(R_{j})\left(\begin{array}{c}
\Upsilon_{\ell,j-1}\\
\Lambda_{\ell,j-1}
\end{array}\right),
\label{Eq:System_bound_cond}
\end{equation}
where
\begin{equation}
\mathbb{D}_{\ell,j}(r) := 
\left(\begin{array}{cc}
X_{\nearrow\ell,j}(r) & X_{\nwarrow\ell,j}(r) \\
\frac{dX_{\nearrow\ell,j}}{dr_{*}}(r) & \frac{dX_{\nwarrow\ell,j}}{dr_{*}}(r)
\end{array}\right).
\label{Eq:Matrix_D_ell_O}
\end{equation}
The determinant of the matrix~\eqref{Eq:Matrix_D_ell_O}  is independent of $r$, since both $e^{st} X_{\nearrow\ell,j}(r)$ and $e^{st} X_{\nwarrow\ell,j}(r)$ are solutions of the RW equation~\eqref{Eq:ReggeWheeler}. Evaluating the determinant (for fixed mass) in the limit $r\to\infty$ one finds
\begin{equation}
\det\left(\mathbb{D}_{\ell,j}(r)\right) = 2s,
\label{app_det}
\end{equation}
and thus the matrix is invertible as long as $s = -i\omega\neq 0$.

Finally, for future reference, we write the explicit form of the tortoise coordinate $r_*$ at each shell. Using Eq.~\eqref{Eq:explicit_tortoise_coo}, one obtains
\begin{equation}
r_*(R_j) = R_j +
2\sum_{k=1}^{j-1}m_{k}\log{\left(
\frac{R_{k+1}-2m_k}{R_k-2m_k}
\right)},
\label{Eq:tortoise_coo_eachR}
\end{equation}
for $j=1,2,\ldots,N$.

\section{Transfer matrix method}
\label{Sec:TransferMatrix}

In this section we introduce the transfer matrix formalism, and we compute the explicit form of the transfer matrices using a perturbative approach. To this purpose it is convenient to rewrite Eq.~\eqref{Eq:System_bound_cond} in the form
\begin{equation}
\left(\begin{array}{c}
\Upsilon_{\ell,j}\\
\Lambda_{\ell,j}
\end{array}\right) = 
\mathbb{M}_j\left(\begin{array}{c}
\Upsilon_{\ell,j-1}\\
\Lambda_{\ell,j-1}
\end{array}\right)	,
\label{Eq:Amplitudes_iteration}
\end{equation}
where we have introduced the transfer matrix
\begin{equation}
\mathbb{M}_j := \mathbb{D}_{\ell,j}(R_j)^{-1}\mathbb{D}_{\ell,j-1} (R_j).
\label{transf_mat}
\end{equation}
The transfer matrix~\eqref{transf_mat} matches the amplitudes of the in- and out-going waves on both sides of the $j$-th shell. The explicit form of this matrix is
\begin{equation}
\mathbb{M}_j = \left( \begin{array}{ll} 
(\mathbb{M}_{j})_{11} & (\mathbb{M}_{j})_{12} \\
(\mathbb{M}_{j})_{21} & (\mathbb{M}_{j})_{22}
\end{array}
\right).
\end{equation}
with
\begin{eqnarray}
(\mathbb{M}_{j})_{11} &=&
\frac{1}{2s}\left( \frac{dX_{\nwarrow\ell,j}}{dr_*} X_{\nearrow\ell,j-1} 
- X_{\nwarrow\ell,j}\frac{dX_{\nearrow\ell,j}}{dr_*} \right),\qquad\\
(\mathbb{M}_{j})_{12} &=&
\frac{1}{2s}\left( \frac{dX_{\nwarrow\ell,j}}{dr_*} X_{\nwarrow\ell,j-1} 
- X_{\nwarrow\ell,j}\frac{dX_{\nwarrow\ell,j}}{dr_*} \right),\qquad\\
(\mathbb{M}_{j})_{21} &=&
\frac{1}{2s}\left( -\frac{dX_{\nearrow\ell,j}}{dr_*} X_{\nearrow\ell,j-1} 
+ X_{\nearrow\ell,j}\frac{dX_{\nearrow\ell,j}}{dr_*} \right),\qquad\\
(\mathbb{M}_{j})_{22} &=&
\frac{1}{2s}\left( -\frac{dX_{\nearrow\ell,j}}{dr_*} X_{\nwarrow\ell,j-1} 
+ X_{\nearrow\ell,j}\frac{dX_{\nwarrow\ell,j}}{dr_*} \right),\qquad
\end{eqnarray}
where it is understood that the expressions on the right-hand side are evaluated at $r = R_j$.

Taking into account Eq.~(\ref{Eq:XnenwSym}) and the fact that $s = -i\omega$ is purely imaginary it is easy to see that
\begin{align}
(\mathbb{M}_{j})_{22} &= \overline{(\mathbb{M}_{j})_{11}}, 
\label{Eq:M22_tM22} \\
(\mathbb{M}_{j})_{21} &= \overline{(\mathbb{M}_{j})_{12}}.
\label{Eq:M21_tM21}
\end{align}
From Eq.~\eqref{app_det} it also follows immediately that the determinant of
$\mathbb{M}_j(r)$ is unitary
\begin{equation}
\det(\mathbb{M}_j) = 1.
\label{Eq:Det_TransMat}
\end{equation}

By multiplying successive transfer matrices, one obtains the total transfer matrix
\begin{equation}
\mathbb{M}_T := \mathbb{M}_N\mathbb{M}_{N-1}\cdots\mathbb{M}_1
\label{Eq:Transfer_matrix_T},
\end{equation}
which connects the amplitudes $\Upsilon_{\ell,0}$ and $\Lambda_{\ell,0}$ in the inner Minkowski space with the amplitudes $\Upsilon_{\ell,N}$ and $\Lambda_{\ell,N}$ of the waves outside of the $N$-th shell
\begin{equation}
\left(\begin{array}{c}
\Upsilon_{\ell,N}\\
\Lambda_{\ell,N}
\end{array}\right) = 
\mathbb{M}_T\left(\begin{array}{c}
\Upsilon_{\ell,0}\\
\Lambda_{\ell,0}
\end{array}\right)	.
\label{Eq:Ampli_ft_M_T}
\end{equation}
Note that the total transfer matrix~(\ref{Eq:Transfer_matrix_T}) also satisfies the properties \eqref{Eq:M22_tM22}, \eqref{Eq:M21_tM21}, and \eqref{Eq:Det_TransMat}.

\subsection{Perturbative expansion of transfer matrices}
\label{Sec:PertTransferMatrix}

It is useful to write the matrix $\mathbb{D}_{\ell,j}$ as the sum of two terms
\begin{equation}
\mathbb{D}_{\ell,j}(r) = \mathbb{D}^{(0)}_{\ell}(r)
+ \frac{2m_{j}}{r}\frac{\mathbb{D}^{(1)}_{\ell}(r)}{\omega r},
\label{d_split}
\end{equation}
where the first matrix on the right-hand side of Eq.~(\ref{d_split}) does not depend on the mass function $m(r)$, while the second summand is proportional to the weak-field term $2m(r)/r$. After introducing the matrix
\begin{equation}
\mathbb{B}_{\ell}(r) := 
[\mathbb{D}^{(0)}_{\ell}(r)]^{-1}\mathbb{D}^{(1)}_{\ell}(r),
\label{Eq:Matrix_B_ell}
\end{equation}
one can factor out $\mathbb{D}^{(0)}_{\ell}(r)$ in Eq.~(\ref{d_split}) and write
\begin{equation}
\mathbb{D}_{\ell,j}(r) = \mathbb{D}^{(0)}_{\ell}(r)\left(\mathds{1}+\frac{2m_{j}}{r}\frac{\mathbb{B}_{\ell}(r)}{\omega r}\right)	.
\label{Eq:Matrix_D_ell}
\end{equation}
This makes possible to express the transfer matrix~(\ref{transf_mat}) as
\begin{equation}
\mathbb{M}_j = \left[ \mathds{1} + \frac{2m_{j}}{R_j}\frac{\mathbb{B}_{\ell}(R_j)}{\omega R_j} \right]^{-1} 
    \left[ \mathds{1} + \frac{2m_{j-1}}{R_j}\frac{\mathbb{B}_{\ell}(R_j)}{\omega R_j} \right] .
\label{m_j}
\end{equation}

It is important to observe that the matrix~(\ref{Eq:Matrix_B_ell}) is bounded for finite values of $\omega r$. It remains bounded in the limit $\omega r\to \infty$ as can be seen from the high-frequency asymptotic expansion
\begin{equation}
	\mathbb{B}_\ell (r) = \mathbb{B}^{(0)}_\ell (r) +
	\frac{1}{\omega r} \mathbb{B}^{(1)}_\ell (r) + 
	\mathcal{O}\left(\frac{1}{(\omega r)^{2}} \right)
	\label{b_expansion}
\end{equation}
with
\begin{equation}
	\mathbb{B}^{(0)}_\ell (r) = \frac{1-S^2}{4}
	\left(\begin{array}{cc}
		i & 0\\
		0 & -i
	\end{array}\right)
	\label{b_0}
\end{equation}
and
\begin{equation}
	\mathbb{B}^{(1)}_\ell (r) = 
	- \frac{\ell(\ell +1)-1+S^2}{4}
	\left(\begin{array}{cc}
		0 & e^{-2i\omega r_{*}}	\\
		e^{2i\omega r_{*}} & 0 	
	\end{array}\right).
	\label{b_1}
\end{equation}
Note that the zeroth order term~(\ref{b_0}) gives the dominant
contribution in the high-frequency limit for scalar ($S = 0$) and
gravitational fields ($S = 2$), whereas it vanishes for an
electromagnetic field ($S = 1$). In the latter case the
first-order term, proportional to the matrix~(\ref{b_1}),
provides the leading contribution.

The fact that the matrix~(\ref{Eq:Matrix_B_ell}) is bounded allows one to expand the right-hand side of Eq.~(\ref{m_j}) in the weak field limit~\eqref{Eq:WeakFieldLimit} and approximate the single-shell transfer
matrix as
\begin{equation}
\mathbb{M}_j = \mathds{1}-\frac{2\Delta m_{j}}{\omega R_{j}^2}\mathbb{B}_{\ell}(R_{j})
+\mathcal{O}\left(\left(\frac{2m_{j}}{\omega R_{j}^2}\mathbb{B}_{\ell}(R_{j})\right)^{2}\right)	,
\label{m_close_to_1}
\end{equation}
where the matrix $\mathbb{B}_{\ell}(R_{j})$ must be understood
as a short-hand notation for the truncated expansion~(\ref{b_expansion}).

Eq.~(\ref{m_close_to_1}) shows that the
single-shell transfer matrices share two features: a) they are
``close'' to the unit matrix, i.e., they have the form
\begin{equation}
	\mathbb{M}_j = \mathds{1} -
	\epsilon_{j} \mathbb{B}_{\ell}(R_{j}) + \ldots ,
\end{equation}
and b) their ``distance'' from the unit matrix, which is measured
by the parameter $\epsilon_{j}$, decreases as $1/j^2$. As discussed in
Appendix~\ref{App:TranM_T_error}, these two features ensure that the total transfer
matrix~(\ref{Eq:Transfer_matrix_T}) can be written in the form
\begin{multline}
\mathbb{M}_T = \mathds{1} -
2\sum_{k=1}^{N}\frac{\Delta m}{\omega (R_1+(k-1)\Delta R)^2}\mathbb{B}_{\ell}(R_{k}) \\ +
\mathcal{O}\left(\left(\frac{\Delta m}{\omega(\Delta R)^2}\right)^{2}\right).
\label{Eq:Transfer_matrix_T_E}
\end{multline}
We remark that condition a), in itself, does not guarantee
the validity of the expansion~(\ref{Eq:Transfer_matrix_T_E});
it is also necessary that the parameters $\epsilon_{j}$ decrease
sufficiently fast for increasing values of $j$. In the present
case both conditions are fulfilled.

\section{Transmission and reflection coefficients}
\label{Sec:TR}

We now analyze the transport properties of the spacetime defined in Sec.~\ref{Sec:Spacetime}. We focus our attention on the propagation of a monochromatic spherical wave of frequency $\omega$ and total angular momentum $\ell$. We consider a wave which, after being radiated from the center of the inner Minkowski space, crosses $N$ thin concentric shells. Our purpose is to compute the corresponding transmission and reflection coefficients.

To achieve this goal, we first have to determine the energy flux associated with a spherical wave. In the case of scalar ($S = 0$) and electromagnetic ($S = 1$) waves, incoming and outgoing fluxes can be computed in a straightforward way by means of the energy-momentum tensor associated with the field. The case of linearized gravitational waves ($S = 2$) is more delicate, because for gravitational radiation there is not an energy-momentum tensor with the properties of being covariantly defined, local, and divergence-free (see, for instance, Refs.~\cite{book:Straumann2013}). Nevertheless, as discussed below, the covariant interpretation of the RW equation~\eqref{Eq:ReggeWheeler}~\cite{Art:Gauge2001Sarbach,Art:linearpert2013Chaverra} allows one to introduce a universal \emph{effective} energy-momentum tensor $T_{\mu\nu}$ which gives rise to a conserved current when contracted with the time-like Killing vector field of the Schwarzschild spacetime, even if the tensor $T_{\mu \nu}$ has a non-vanishing divergence. In this way it is possible to compute the desired conserved flux for fields of arbitrary spin $S$. For $S=0$, the effective $T_{\mu\nu}$ coincides with the usual energy-momentum tensor of the scalar field (which is divergence-free), whereas for $S=1$ it is shown in Appendix~\ref{App:EM_flux} that the resulting fluxes agree (up to a constant factor) with those associated with the Poynting vector.

Our universal approach is based on rewriting the RW equation~\eqref{Eq:ReggeWheeler} in the covariant form~\cite{Art:linearpert2013Chaverra,Art:Cauchy2018Ortiz}
\begin{equation}
\square\Psi + V(r)\Psi = 0,\qquad V(r) = -\frac{2m(r)S^2}{r^3},
\label{Eq:ReggeWheeler_scalarForm}
\end{equation}
where $\square := -\nabla^\mu\nabla_\mu$ is the curved spacetime d'Alembert operator and $\Psi$ is a (real-valued) scalar function which admits the series representation
\begin{equation}
\Psi(t,r,\vartheta,\varphi) 
 = \frac{1}{r}\sum\limits_{\ell=S}^\infty\sum\limits_{m=-\ell}^\ell \Phi_{\ell m}(t,r) Y^{\ell m}(\vartheta,\varphi) .
\label{Eq:PsiExpansion}
\end{equation}
In Eq.~(\ref{Eq:PsiExpansion}) the functions $\Phi_{\ell m}$ are solutions of the RW equation~\eqref{Eq:ReggeWheeler}, while the functions $Y^{\ell m}$ are standard spherical harmonics, and we have momentarily reintroduced the magnetic quantum number $m$ which should not be confused with the mass function $m(r)$. The functions $\Phi_{\ell m}$ should be chosen such that $\overline{\Phi_{\ell m}} = (-1)^m\Phi_{\ell,-m}$ in order to guarantee that $\Psi$ is real-valued. Equation~\eqref{Eq:ReggeWheeler_scalarForm} describes a scalar field $\Psi$ which is subject to the external effective potential $V(r)$. Note that for $S=0$ this potential vanishes and Eq.~(\ref{Eq:ReggeWheeler_scalarForm}) reduces to the standard wave equation.

Motivated by these observations, we introduce the effective energy-momentum tensor
\begin{equation}
T_{\mu\nu} := (\nabla_{\mu}\Psi)(\nabla_{\nu}\Psi) - \frac{1}{2}g_{\mu\nu}\left[ (\nabla^{\alpha}\Psi)(\nabla_{\alpha}\Psi) + V(r)\Psi^2\right]
\label{Eq:ene-mom-tensor_wfield}
\end{equation}
which is symmetric by definition and, according to Eq.~\eqref{Eq:ReggeWheeler_scalarForm}, satisfies the divergence law
\begin{equation}
\nabla^{\mu}T_{\mu\nu} = -\frac{1}{2}\frac{dV}{dr}(r)(\nabla_{\nu}r)\Psi^2.
\label{Eq:DivT}
\end{equation}
As stated above, the right-hand side is only zero in the scalar case, i.e., when $S=0$. However, when both sides of Eq.~(\ref{Eq:DivT}) are contracted with the time-like Killing vector field $k := \partial/\partial t$, one obtains the conservation law
\begin{equation}
\nabla_{\mu}J^{\mu}_{\epsilon} = 0,\qquad
J^{\mu}_{\epsilon} := -T^{\mu}{_{\nu}}k^{\nu},
\label{Eq:conserved_current}
\end{equation}
since $k[r] = k^\nu\nabla_\nu r = 0$. As shown below, the presence of this conserved current allows one to compute the in- and outflow of scalar, electromagnetic and linearized ``gravitational" energy across surfaces with a fixed radius over a given time period in a fully covariant way. This finding underscores that the presence of a divergence-free energy-momentum tensor is not a prerequisite for defining conserved currents. Nonetheless, the physical nature of the effective tensor~\eqref{Eq:ene-mom-tensor_wfield} remains elusive.

\subsection{Energy flux through a sphere}
\label{flux_sphere}

The energy that is radiated through a sphere $S_r^2$ of constant radius $r$ during the time interval $\Delta t$ is given by the flux integral
\begin{equation}
\Delta F = \int\displaylimits_{[t,t+\Delta t]\times S_r^2} J^\mu_\epsilon n_{\mu}d\Sigma,
\label{Delta_F}
\end{equation}
over the three-dimensional hypersurface $[t,t+\Delta t]\times S_r^2$. In Eq.~(\ref{Delta_F}) $d\Sigma$ denotes a surface element and $n_\mu$ is the unit outward co-vector to this surface. For the spacetime with metric~\eqref{Eq:metric_full_st} one has
\begin{equation}
n_{\mu}=\frac{\nabla_{\mu}r}{\sqrt{1-\frac{2m(r)}{r}}}, \qquad
d\Sigma=\sqrt{1-\frac{2m(r)}{r}}r^{2}dtd\Omega,
\end{equation}
and hence the average energy radiated per unit time is given by
\begin{equation}
\frac{\Delta F}{\Delta t} = \frac{r^2}{\Delta t}\int\limits_t^{t + \Delta t} dt\int\limits_{S_r^2} d\Omega J_{\epsilon}^r.
\label{Eq:flux_PUT}
\end{equation}
Note that this average only depends on the radial component of the current, which is equal to
\begin{equation}
J_{\epsilon}^r = -T^r{}_t 
 = -\left( \frac{\partial\Psi}{\partial r_*} \right)\left( \frac{\partial\Psi}{\partial t} \right).
\label{Eq:radial_current}
\end{equation}
We remind the reader that the tortoise coordinate $r_*$ is defined by Eq.~\eqref{Eq:tortoise_coordinate}. We next introduce the expansion~\eqref{Eq:PsiExpansion} in Eq.~(\ref{Eq:radial_current})
and integrate the radial current over a sphere. With the help of the orthonormality relations of the spherical harmonics, we obtain
\begin{align}
& r^2\int\limits_{S_r^2} d\Omega J_{\epsilon}^r 
 = -\sum\limits_{\ell=S}^\infty\sum\limits_{m=-\ell}^\ell 
 \Re\left[ \left( \frac{\partial\overline{\Phi_{\ell m}}}{\partial r_*} \right)
 \left( \frac{\partial\Phi_{\ell m}}{\partial t} \right) \right]
\nonumber\\
 &+ \frac{1}{r}\left( 1 - \frac{2m(r)}{r} \right) 
 \sum\limits_{\ell=S}^\infty\sum\limits_{m=-\ell}^\ell 
 \Re\left( \overline{\Phi_{\ell m}} \frac{\partial\Phi_{\ell m}}{\partial t} \right).
 \label{twoterms}
\end{align}
Taking into account the monochromatic ansatz~\eqref{Eq:MonochromaticInOut}
and the identity~\eqref{Eq:XnenwSym}, we conclude that the second series in the right-hand side of Eq.~\eqref{twoterms} vanishes, while the summands in the first series are equal to
\begin{equation}
\frac{i\omega}{2}\left( |\Lambda_{\ell m}|^2 - |\Upsilon_{\ell m}|^2 \right)
\det\left( \begin{array}{cc}
 \displaystyle 	X_{\nearrow,\ell m} & \displaystyle X_{\nwarrow,\ell m} \\
 \displaystyle \frac{\partial X_{\nearrow,\ell m}}{\partial r_*} &
 \displaystyle \frac{\partial X_{\nwarrow,\ell m}}{\partial r_*}
 \end{array}
\right).
\label{wron}
\end{equation}
The Wronskian in the previous expression is equal to the determinant of the
matrix~(\ref{Eq:Matrix_D_ell_O}), i.e., to $2s = -2i\omega$. One is thus led
to the conclusion that the average energy flux through a sphere of constant radius $r$ is
\begin{equation}
\frac{\Delta F}{\Delta t} 
 = \omega^2\sum\limits_{\ell=S}^\infty\sum\limits_{m=-\ell}^\ell 
\left( |\Upsilon_{\ell m}|^2  - |\Lambda_{\ell m}|^2 \right) .
\label{flux}
\end{equation}
Note that the flux~(\ref{flux}) does not depend on the surface radius $r$; it is
proportional to the square of frequency $\omega$, and positive for purely outgoing waves and negative for incoming ones.

As an additional remark, we observe that Eq.~(\ref{flux}) makes possible to prove that the conservation law~(\ref{Eq:conserved_current}) is consistent with the unimodularity of the transfer matrices, see Eq.~(\ref{Eq:Det_TransMat}).
In fact, when the condition of flux
conservation~(\ref{Eq:conserved_current}) is applied
to a monochromatic wave crossing the $j$-th shell, one
obtains
\begin{equation}
	|\Upsilon_{\ell m,j}|^{2} - |\Lambda_{\ell m,j}|^{2} =
	|\Upsilon_{\ell m,j-1}|^{2} - |\Lambda_{\ell m,j-1}|^{2} .
	\label{conserved_flux}
\end{equation}
With the help of Eq.~(\ref{Eq:Amplitudes_iteration}),
one can write the previous identity as
\begin{equation}
\begin{array}{cl}
     & 	\left[ |(\mathbb{M}_{j})_{11}|^{2} - |(\mathbb{M}_{j})_{12}|^{2} \right]
	\left[ |\Upsilon_{\ell m,j-1}|^{2} - |\Lambda_{\ell m,j-1}|^{2} \right] \\
	= & |\Upsilon_{\ell m,j-1}|^{2} - |\Lambda_{\ell m,j-1}|^{2} ,
\end{array}	
\end{equation}
which leads to the conclusion that
\begin{equation}
	|(\mathbb{M}_{j})_{11}|^{2} - |(\mathbb{M}_{j})_{12}|^{2} = 1 .
	\label{unimod}
\end{equation}
Eq.~(\ref{unimod}), together with conditions~(\ref{Eq:M22_tM22})
and~(\ref{Eq:M21_tM21}), implies that the determinant of the
transfer matrix $\mathbb{M}_{j}$ is equal to one.

\subsection{Transmission and reflection coefficients}

The transmission and reflection coefficients across $N$ shells are defined as follows:
\begin{align}
    \mathscr{T} &:= \frac{\text{Transmitted flux}}{\text{Incident flux}},
    \label{Eq:T_coeff} \\
    \mathscr{R} &:= \frac{|\text{Reflected flux}|}{\text{Incident flux}}.
    \label{Eq:R_coeff}
\end{align}
Taking into account Eq.~(\ref{flux}), for a monochromatic
wave of angular momentum $\ell$ and magnetic quantum number $m$, these coefficients become
\begin{equation}
\mathscr{T} = \frac{|\Upsilon_{\ell m,N}|^{2}}{|\Upsilon_{\ell m,0}|^{2}},
\quad
\mathscr{R} = \frac{|\Lambda_{\ell m,0}|^{2}}{|\Upsilon_{\ell m,0}|^{2}}.
\label{Eq:TR}
\end{equation}
Equation~(\ref{Eq:Ampli_ft_M_T}) shows that the amplitudes
of the impinging, reflected, and transmitted waves are linked by the
total transfer matrix. Assuming that no incoming radiation reaches
the outer $N$-th shell from the external region, Eq.~\eqref{Eq:Ampli_ft_M_T}
can be written as
\begin{align}
    0 &= (M_{T})_{11}\Lambda_{\ell m,0}+(M_{T})_{12}\Upsilon_{\ell m,0}, \\
    \Upsilon_{\ell m,N} &= (M_{T})_{21}\Lambda_{\ell m,0}+(M_{T})_{22}\Upsilon_{\ell m,0}.
\end{align}
These equations, together with the properties~(\ref{Eq:M22_tM22}), (\ref{Eq:M21_tM21}), and~(\ref{Eq:Det_TransMat}) of the total
transfer matrix, make possible to write the transmission and reflection coefficients as
\begin{equation}
    \mathscr{R} =\frac{|(M_{T})_{12}|^{2}}{|(M_{T})_{11}|^{2}},\quad
    \mathscr{T} =\frac{1}{|(M_{T})_{11}|^{2}} .
\label{Eq:RyT_Coeff}
\end{equation}

Note that the condition of unitary determinant implies that
\begin{equation}
	 \mathscr{T} + \mathscr{R} = 1,
	 \label{Eq:T+R}
\end{equation}
as expected.

\section{Results}
\label{Sec:Results}

In this section we present the most relevant results for the reflection and transmission coefficients in the weak field limit. We provide explicit expressions for $\mathscr{R}$ and $\mathscr{T}$ for scalar, electromagnetic and odd-parity linearized gravitational radiation fields with total angular momentum number $\ell$ satisfying $S\leq \ell\leq 4$. Furthermore, we analyze in greater detail the transmission properties of our model in the high-frequency limit.

To compute the transmission and reflection coefficients one must insert the elements of the transfer matrix~\eqref{Eq:Transfer_matrix_T_E} into Eq.~\eqref{Eq:RyT_Coeff}, which yields
\begin{eqnarray}
\mathscr{T} &=& \frac{1}{\left|1-\frac{2\Delta m}{\omega}\sum_{k=1}^{N}\frac{\left(\mathbb{B}_{\ell}(R_{k})\right)_{11}}{\left(R_{1}+(k-1)\Delta R\right)^2}\right|^{2}} \nonumber\\
 &+& \mathcal{O}\left(\left(\frac{\Delta m}{\omega \Delta R^2}\right)^2\right),
\label{Eq:TExpl}\\
\mathscr{R} &=& \frac{
\left|\frac{2\Delta m}{\omega}\sum_{k=1}^{N}\frac{\left(\mathbb{B}_{\ell}(R_{k})\right)_{12}}{\left(R_{1}+(k-1)(\Delta R)\right)^2}\right|^2}
{\left|1-\frac{2\Delta m}{\omega}\sum_{k=1}^{N}\frac{\left(\mathbb{B}_{\ell}(R_{k})\right)_{11}}{\left(R_{1}+(k-1)\Delta R\right)^2}\right|^2} \nonumber\\
 &+& \mathcal{O}\left(\left(\frac{\Delta m}{\omega(\Delta R)^2}\right)^3\right).
\label{Eq:RExpl}
\end{eqnarray}
It is important to observe that the matrices $\mathbb{B}_{\ell}(R_{k})$ have purely imaginary diagonal elements, as can be deduced from the properties of the matrices $\mathbb{D}_{\ell,j}(r)$, see Appendix~\ref{App:Aux_M}. As a consequence, the denominators on the right-hand sides of Eqs.~\eqref{Eq:TExpl} and~\eqref{Eq:RExpl} differ from one by a term which is quadratic rather than linear in the expansion parameter $\Delta m/(\omega(\Delta R)^2)$. Since we have determined $\mathbb{B}_{\ell}(R_{k})$ only up to first-order terms, we cannot use Eq.~\eqref{Eq:TExpl} to determine the leading-order correction of the transmission coefficient. However, we can use Eq.~\eqref{Eq:RExpl} to compute  the reflection coefficient in the second-order approximation and then obtain $\mathscr{T}$ from the property~\eqref{Eq:T+R} of flux conservation.

\subsection{Weak field reflection coefficient}
\label{Sub:WF_RCoeff}

Using Eqs.~(\ref{Eq:RadialW_sol},\ref{Eq:Matrix_D_ell_O},\ref{d_split},\ref{Eq:Matrix_B_ell}), one obtains from Eq.~\eqref{Eq:RExpl} the following explicit expressions for the reflection coefficient, neglecting third or higher order terms in the expansion parameter:
\begin{enumerate}
	\item For the case $\ell=S=0$,
	\begin{equation}
		\mathscr{R}\simeq \left(\frac{\Delta m}{\omega}\right)^{2}
		\left|\sum_{k=1}^{N}\frac{e^{2i\omega\left(R_{k}-r_{*}(R_{k})\right)}E_{3}(2i\omega R_{k})}{\left(R_{1}
			+(k-1)\Delta R\right)^{2}}\right|^{2}	.
		\label{Eq:WF_R_l0}
	\end{equation}
	\item For the case $\ell=1$ and $S=0$ or $S=1$:
		\begin{multline}
			\mathscr{R} \simeq \left(\frac{\Delta m}{\omega}\right)^{2}
			\Biggl|\sum_{k=1}^{N}\frac{e^{-2i\omega r_{*}(R_{k})}}{\left(R_{1}
				+(k-1)\Delta R\right)^{2}}  \\ \times 
			\left[\frac{S^{2}}{4\left(\omega R_{k}\right)^{2}}\left(i-2\omega R_{k}
			\right)-ie^{2i\omega R_{k}}E_{3}(2i\omega R_{k})\right] \Biggr|^{2}		.
			\label{Eq:WF_R_l1}
		\end{multline}
	\item For the case $\ell = 2$ and $S=0,1$ or $2$,
		\begin{multline}
			\mathscr{R} \simeq \left(\frac{\Delta m}{\omega}\right)^{2}
			\Biggl|\sum_{k=1}^{N}\frac{e^{-2i\omega r_{*}(R_{k})}}{\left(R_{1}
			+(k-1)\Delta R\right)^{2}}  \\ \times 
			\Biggl[\frac{3\left(S^{2}
			+3\right)}{2\left(\omega R_{k}\right)^{4}}\left(2\omega R_{k}-i\right)
			+\frac{3\left(S^{2}+10\right)i}{4\left(\omega R_{k}\right)^{2}} \\ 
			-\frac{S^{2}+6}{2\omega R_{k}}
			+ie^{2i\omega R_{k}}E_{3}\left(2i\omega R_{k}\right)\Biggr]\Biggr|^{2}	.
			\label{Eq:WF_R_l2}
		\end{multline}
	\item For the case $\ell = 3$ and $S=0,1$ or $2$,
		\begin{multline}
			\mathscr{R} \simeq \left(\frac{\Delta m}{\omega}\right)^{2}
			\Biggl|\sum_{k=1}^{N}\frac{e^{-2i\omega r_{*}(R_{k})}}{\left(R_{1}
				+(k-1)\Delta R\right)^{2}}  \\ \times
			\Biggl[\frac{225(S^{2}+8)}{8(\omega R_{k})^{6}}\left(i - 2\omega R_{k}\right) 
			-\frac{15(13S^{2}+106)i}{4(\omega R_{k})^{4}} \\
			+\frac{15(3S^{2}+26)}{2(\omega R_{k})^{3}}
			+\frac{21(S^{2}+10)i}{4(\omega R_{k})^{2}} \\
			-\frac{S^{2}+10}{2\omega R_{k}}
			-ie^{2i\omega R_{j}}E_3(2i\omega R_k)\Biggr]\Biggr|^{2}	.
			\label{Eq:WF_R_l3}
		\end{multline}
	\item For the case $\ell = 4$ and $S=0,1$ or $2$,
		\begin{multline}
			\mathscr{R}\simeq \left(\frac{\Delta m}{\omega}\right)^{2}
			\Biggl|\sum_{k=1}^{N}\frac{e^{-2i\omega r_{*}(R_{k})}}{\left(R_{1}
				+(k-1)\Delta R\right)^{2}}   \\ \times 
			\Biggl[\frac{2205(S^{2}+15)}{2(\omega R_{k})^{8}}\left(2\omega R_{k}-i\right)
			+\frac{315(51S^{2}+770)i}{8(\omega R_{k})^{6}} \\
			-\frac{105(41S^{2}+630)}{4(\omega R_{k})^{5}}
			-\frac{15(97S^{2}+1544)i}{4(\omega R_{k})^{4}} \\ 
			+\frac{3(51S^{2}+860)}{2(\omega R_{k})^{3}}
			+\frac{37S^{2}+660}{4(\omega R_{k})^{2}}i \\
			-\frac{S^{2}+20}{2\omega R_{k}}
			+ie^{2i\omega R_{k}}E_{3}(2i\omega R_{k})\Biggr]\Biggr|^2.
			\label{Eq:WF_R_l4}
		\end{multline}
\end{enumerate}
In these formulae the exponential term $e^{-2i\omega r_*(R_k)}$ can be computed with the help of Eq.~(\ref{Eq:tortoise_coo_eachR}), which yields
\begin{multline}
	\exp\left( -2i\omega r_*(R_k) \right) = 
	\exp[-2i\omega R_1 
	\left(1+(k-1)\frac{\Delta R}{R_1}\right)] \\ 
\times\exp[-4i\omega\Delta m \sum_{j=1}^{k-1} j 
	\log\left(\frac{1 + j \left(\frac{\Delta R}{R_1} - \frac{2\Delta m}{R_1} \right)}
	{1 - \frac{\Delta R}{R_1} + j \left(\frac{\Delta R}{R_1} - \frac{2\Delta m}{R_1}\right)}\right)]  .
\end{multline}
Interestingly, for scalar waves ($S=0$), the reflection coefficients for monopolar ($\ell=0$) and dipolar ($\ell=1$) waves are identical to each other.

\subsection{High frequency reflection coefficient}

The analysis of the behavior of the reflection coefficient in the high-frequency limit (HFL) is particularly significant for several reasons. First, the outcomes observed under the HFL are consistent with those predicted by the Wentzel-Kramers-Brillouin (WKB) method worked out in Appendix~\ref{App:WKB}. This consistency underlines the reliability of the HFL approach in our analysis. Second, the HFL proves to be a valuable tool for understanding the behavior of both transmission and reflection across various types of radiation discussed in this work, irrespective of their total angular momentum. To illustrate the utility of the HFL further, consider the following equation, which encapsulates the derived relationships and quantifies the reflection coefficients in a concise mathematical form:
\begin{equation}
	\mathscr{R} =   F^2
	\left|
	\sum_{k=1}^N \frac{e^{-2i\omega r_*(R_k)}}
	{\left(1+(k-1)\frac{\Delta R}{R_1}\right)^3}
	\right|^2,
	\label{Eq:Ref_coeff_Hfrequency}
\end{equation}
where the factor $F$ is given by
\begin{equation}
F := \frac{\Delta m}{R_1}
	\frac{\ell(\ell+1)-1+S^2}{2(\omega R_1)^2}	.
\label{Eq:R_factor}
\end{equation}
It is remarkable that the dependency on the angular momentum and spin only appears in the factor $F$, which allows us to study the dependency on $\omega$ and $N$ by considering only the second factor in Eq.~(\ref{Eq:Ref_coeff_Hfrequency}). In view of this, it will turn out to be convenient to analyze the quantity $F^{-2}\mathscr{R}$ instead of $\mathscr{R}$, as the former becomes independent of $\ell$ and $S$ in the HFL.

\subsection{Numerical results}

In what follows, we analyze the behavior of the reflection coefficient using the expressions \eqref{Eq:WF_R_l0}, \eqref{Eq:WF_R_l1}, \eqref{Eq:WF_R_l2}, and \eqref{Eq:Ref_coeff_Hfrequency}, as a function of the following dimensionless parameters:
\begin{itemize}
	\item $\omega R_1$ (dimensionless frequency),
	\item $\Delta m/R_1$ (perturbative parameter),
	\item $\Delta R/R_1$ (distance between the shells in units of $R_1$).
\end{itemize}
Note that $\Delta R/R_1$ determines the average mass density of the system (i.e., its total mass divided by its volume), small values corresponding to high average densities and vice-versa. As we will see, this density influences the qualitative behavior of the reflection coefficient. To illustrate this point, we exhibit our results for three different configurations:  low average density ($R_1/\Delta R = 0.2$), medium average density ($R_1/\Delta R = 2$), and high average density ($R_1/\Delta R = 20$). In each case we show results for the reflection coefficient for waves with $S=0,1,2$, and for definiteness we restrict ourselves to $\ell=S$ (however, note that in the HFL, the value of $\ell$ only affects the factor $F$). A direct comparison between the low, medium and high average density systems will also be provided at the end of this section.

\subsubsection{Low average density}

\begin{figure}[htp]
	\includegraphics[clip,width=\columnwidth]{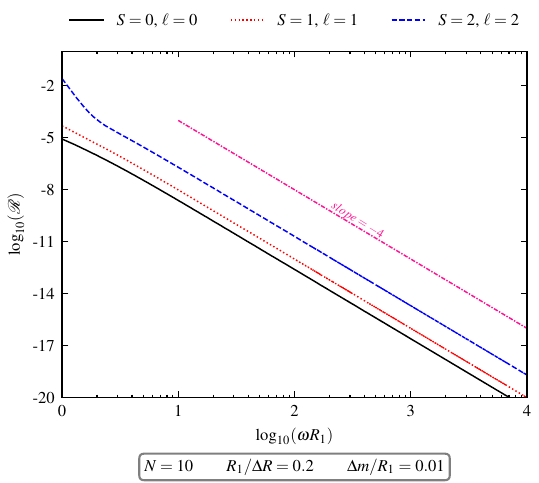}
\caption{Reflection coefficient versus the dimensionless frequency in a log-log scale for different values of $S=\ell$ and the parameter values $N=10$, $R_1/\Delta R = 0.2$, and $\Delta m /R_1 = 0.01$. For reference, we also display a line with slope $-4$ corresponding to the power law decay $(\omega R_1)^{-4}$.}
\label{Fig:ll_FRvsFreq_N10_denLow}
\end{figure}

\begin{figure}[htp]
	\includegraphics[clip,width=\columnwidth]{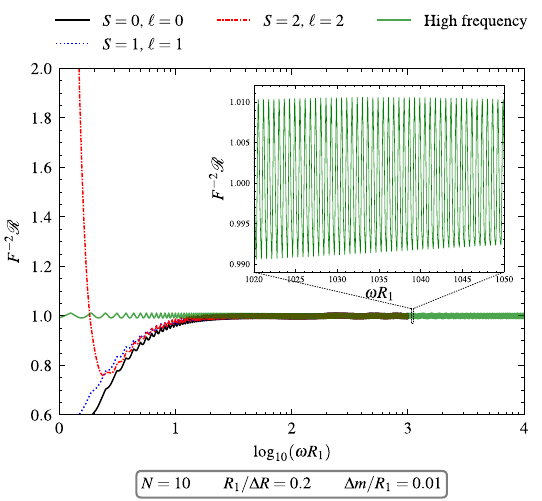}
	\caption{Rescaled reflection coefficient versus the dimensionless frequency in a semilog scale for different values of $S=\ell$ and the parameter values $N=10$, $R_1/\Delta R = 0.2$, $\Delta m/R_1=0.01$. Also shown is the high-frequency expression from Eq.~\eqref{Eq:Ref_coeff_Hfrequency}. The inset represents a zoom in a linear scale over the interval $\omega R_1 \in [1020, 1050]$.}
	\label{Fig:slx_FRvsFreq_N10_denLow}
\end{figure}

Figure~\ref{Fig:ll_FRvsFreq_N10_denLow} shows the reflection coefficient as a function of the  frequency on a log-log scale for $N=10$ shells. We observe that for fixed $\omega$, this coefficient becomes larger as $S$ and $\ell$ increase, as is expected from the behavior of the function $F$ in the HFL. Furthermore, a slope of approximately $-4$ is observed, indicating that the reflection coefficient decays according to the power law $(\omega R_1)^{-4}$. This can again be attributed to the expression for the factor $F$ in the HFL, see Eqs.~\eqref{Eq:Ref_coeff_Hfrequency} and~\eqref{Eq:R_factor}. Figure~\ref{Fig:slx_FRvsFreq_N10_denLow} shows the rescaled reflection coefficient $F^{-2}\mathscr{R}$ as a function of the frequency in a semi-log scale, along with the HFL expression~\eqref{Eq:Ref_coeff_Hfrequency}. As expected, the exact coefficients converge to the ones obtained in the HFL for large values of $\omega R_1$. Note also that the reflection coefficient exhibits an oscillatory behavior in the frequency, albeit with a tiny amplitude.

\begin{figure}[htp]
	\includegraphics[clip,width=\columnwidth]{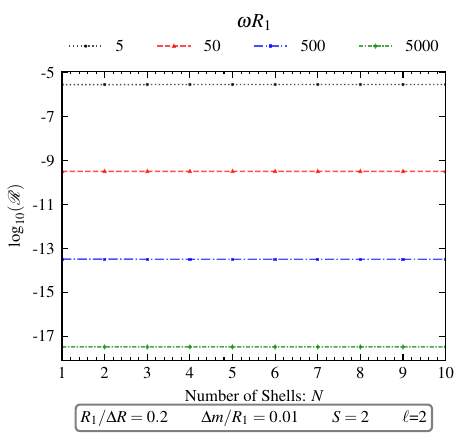}
	\caption{Reflection coefficient versus the number of shells for different values for the dimensionless frequency $\omega R_1$ and the parameters $R_1/\Delta R = 0.2$, $\Delta m/R_1 = 0.01$, $S=2$, and $\ell=2$. The change in the reflection coefficient with respect to $N$ is imperceptible (less than $1\%$), its value saturating from the first shell on. To better illustrate this fact, we have connected the data points at integer values of $N$ with dotted or dashed lines.}
	\label{Fig:RvsN_varFreq_low}
\end{figure}

We have also varied the number of shells $N$ from $1$ to $100$ and found the same qualitative behavior as the one shown in Figs.~\ref{Fig:ll_FRvsFreq_N10_denLow} and~\ref{Fig:slx_FRvsFreq_N10_denLow}, with variations of the reflection coefficient of less than $1\%$. This is further illustrated in Fig.~\ref{Fig:RvsN_varFreq_low}. The fact that for the low average density case the reflection coefficient is nearly independent of $N$ can be understood by looking at the HFL expression~\eqref{Eq:Ref_coeff_Hfrequency}. Indeed, for $\Delta R/R_1 = 5$, the second term in the sum is smaller in magnitude than the first term by a factor of at least $6^3 = 216$.

\subsubsection{Medium average density}

\begin{figure}[htp]
	\includegraphics[clip,width=\columnwidth]{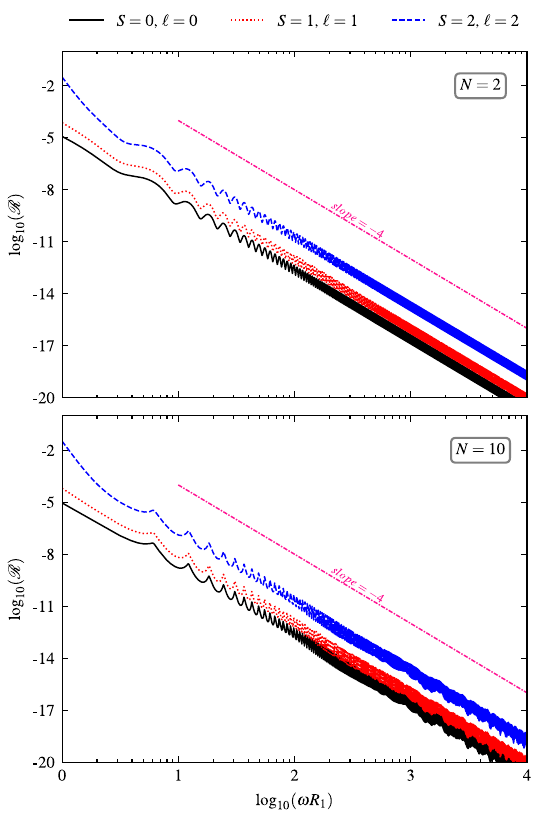}
	\caption{Reflection coefficient versus dimensionless frequency on a log-log scale for different values of $S=\ell$. The parameters are $R_1/\Delta R=2$ and $\Delta m /R_1=0.01$ and $N=2$ (top panel) and $N=10$ (bottom panel).
	}
	\label{Fig:ll_FRvsFreq_N2_10_denMed}
\end{figure}

\begin{figure}[htp]
	\includegraphics[clip,width=\columnwidth]{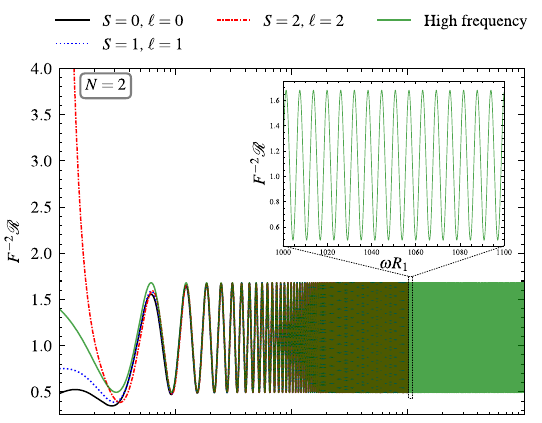} \\
	\includegraphics[clip,width=\columnwidth]{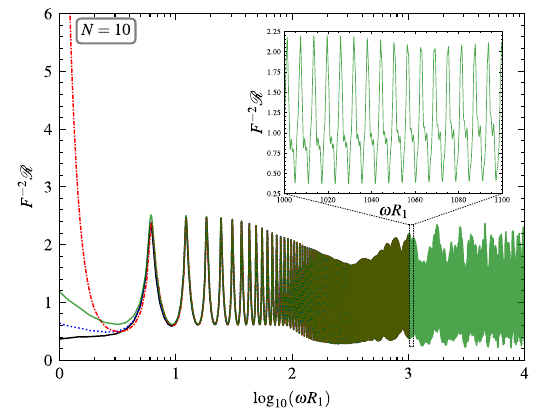 }
	\caption{Rescaled reflection coefficient versus the dimensionless frequency in a semilog scale for different values of $S=\ell$ and the parameter values $R_1/\Delta R = 2$, $\Delta m/R_1=0.01$ and $N=2$ (top panel) or $N=10$ (bottom panel). Also shown is the high-frequency limit obtained from the expression in Eq.~\eqref{Eq:Ref_coeff_Hfrequency}. The insets in both panels provide a zoomed view in a linear scale over the interval $\omega R_1 \in [1000, 1100]$.}
	\label{Fig:slx_FRvsFreq_N2_10_denMed}
\end{figure}

\begin{figure}[htp]
	\includegraphics[clip,width=\columnwidth]{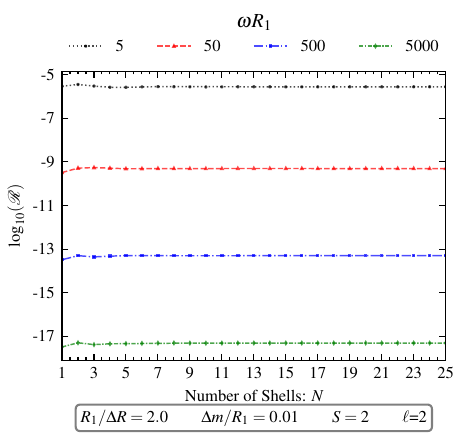}
	\caption{Reflection coefficient versus the number of shells for different values of the dimensionless frequency $\omega R_1$ and the parameters $R_1/\Delta R = 2$, $\Delta m/R_1 = 0.01$, and $S = \ell = 2$. Small fluctuations in the reflection coefficient are observed in the first $5$ shells; however from the sixth shell on these fluctuations are smaller than about $2\%$.}
\label{Fig:RvsN_varFreq_med}
\end{figure}

Figures~\ref{Fig:ll_FRvsFreq_N2_10_denMed}, \ref{Fig:slx_FRvsFreq_N2_10_denMed}, and~\ref{Fig:RvsN_varFreq_med} show the behavior of the reflection coefficient for the medium average density case, and they should be compared with the corresponding figures~\ref{Fig:ll_FRvsFreq_N10_denLow}, \ref{Fig:slx_FRvsFreq_N10_denLow}, and \ref{Fig:RvsN_varFreq_low} for low average density. As can be seen from this comparison, the medium average density reflection coefficient decays again as $(\omega R_1)^{-4}$, as predicted from the factor $F$ in the HFL. However, in contrast to the low average density case, the reflection coefficient exhibits oscillations in the frequency with a much larger amplitude which originate from the sum in the second factor of Eq.~(\ref{Eq:Ref_coeff_Hfrequency}). Furthermore, one observes that these oscillations become more irregular when the number of shells increases from $N=2$ to $N=10$. This effect is particularly visible from the plots in Fig.~\ref{Fig:slx_FRvsFreq_N2_10_denMed} and the inset in the bottom panel of that figure which shows that the oscillations are not sinusoidal.

From Fig.~\ref{Fig:RvsN_varFreq_med} we see that the reflection coefficient shows small fluctuations as a function of $N$ up to the first $6$ shells, after which it saturates to a constant value. As in the previous case, this can be understood by looking at the HFL expression~\eqref{Eq:Ref_coeff_Hfrequency}, for which $\Delta R/R_1 = 1/2$ in the medium average density case, such that the seventh term in the sum is smaller than the first one by a factor of at least $4^3 = 64$, which is consistent with the $2\%$ bound observed for $N\geq 6$.

\subsubsection{High average density}

\begin{figure}[htp]
	\includegraphics[clip,width=\columnwidth]{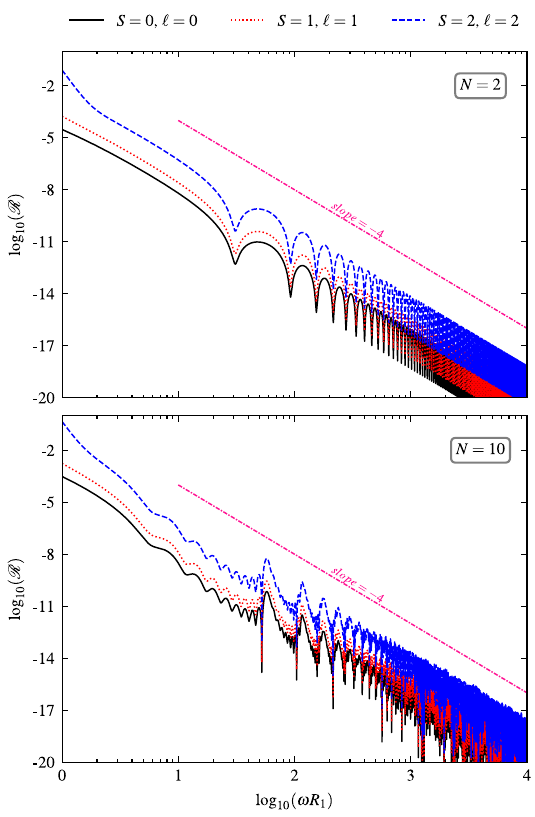}
	\caption{Reflection coefficient versus the dimensionless frequency in a log-log scale for different values of $S=\ell$ and parameters $R_1/\Delta R = 20$, $\Delta m /R_1 = 0.01$ and $N=2$ (top panel) and $N=10$ (bottom panel).}
	\label{Fig:ll_FRvsFreq_N2_10_denHigh}
\end{figure}

\begin{figure}[htp]
	\includegraphics[clip,width=\columnwidth]{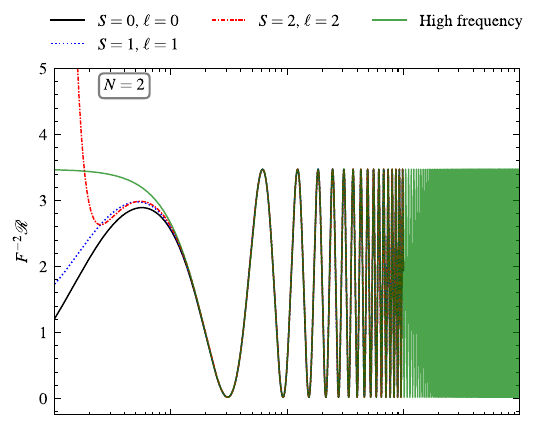}\\
	\includegraphics[clip,width=\columnwidth]{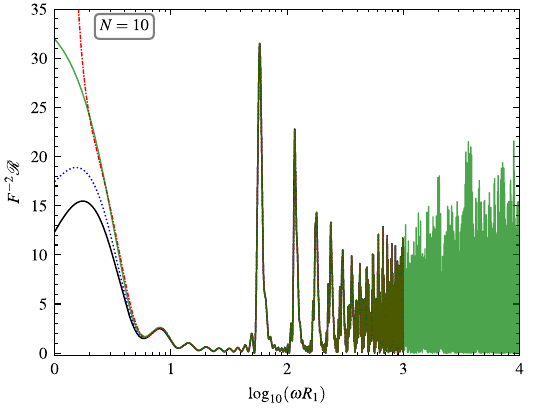}
	\caption{Rescaled reflection coefficient versus the dimensionless frequency in a semilog scale for different values of $S=\ell$ and the parameter values $R_1/\Delta R = 20$, $\Delta m/R_1=0.01$ and $N=2$ (top panel) and $N=10$ (bottom panel). Also shown is the high-frequency limit obtained from the expression in Eq.~\eqref{Eq:Ref_coeff_Hfrequency}.}
	\label{Fig:slx_FRvsFreq_N2_10_denHigh}
\end{figure}

\begin{figure}[htp]
	\includegraphics[clip,width=\columnwidth]{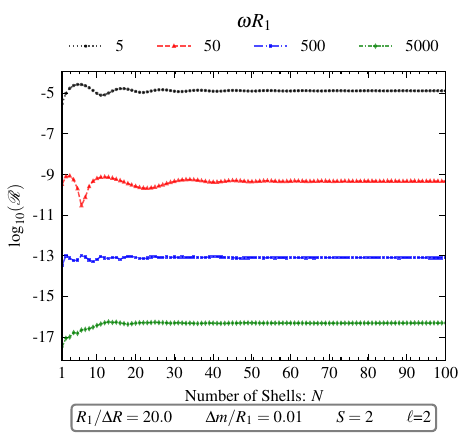}
	\caption{Reflection coefficient versus the number of shells for different values of the dimensionless frequency $\omega R_1$ and the parameters $R_1/\Delta R = 20$, $\Delta m/R_1 = 0.01$, and $S = \ell = 2$. Moderate fluctuations in the reflection coefficient are observed in the first $50$ shells; however from the sixtieth shell on these fluctuations are smaller than about $2\%$.}
	\label{Fig:RvsN_varFreq_denHigh}
\end{figure}

Figures~\ref{Fig:ll_FRvsFreq_N2_10_denHigh}, \ref{Fig:slx_FRvsFreq_N2_10_denHigh}, and~\ref{Fig:RvsN_varFreq_denHigh} show the behavior of the reflection coefficient for a system with high average density. As these figures show, the oscillations' amplitude is even larger than in the medium average density case. They are still regular for $N=2$; however for $N=10$ they present an irregular pattern, as in the medium average density case. Regarding the saturation of the reflection coefficient with the number of shells shown in Fig.~\ref{Fig:RvsN_varFreq_denHigh}, we see that it occurs at about $N=60$. Comparing the sixty-first term in the sum in Eq.~\eqref{Eq:Ref_coeff_Hfrequency} with $\Delta R/R_1 = 1/20$, we obtain a factor of $4^3 = 64$ relative to the first term in the sum, which is consistent with the $2\%$ bound observed for $N\geq 60$.

\subsubsection{Direct comparison between low, medium and high average density}

\begin{figure}[htp]
	\includegraphics[clip,width=\columnwidth]{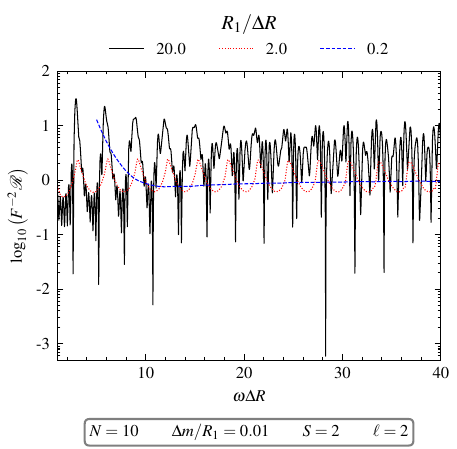}
	\caption{Rescaled reflection coefficient versus the rescaled frequency $\omega\Delta R = \frac{\Delta R}{R_1}\cdot \omega R_1$ in a semilogarithmic scale for fixed parameters $N=10$, $\Delta m/R_1 = 0.01$, $S=2$, and $\ell=2$ and different values of $R_1/\Delta R$.}
	\label{Fig:RvsFreq_varR1oDR}
\end{figure}

Finally, we show in Fig.~\ref{Fig:RvsFreq_varR1oDR} a comparison of the rescaled reflection coefficient $F^{-2}\mathscr{R}$ between the low, medium and high average density systems, as a function of $\omega\Delta R$. As observed previously, in the low density case the reflection coefficient oscillates with a very small amplitude which is not visible in this plot, whereas the oscillations are more pronounced in the medium and high density cases. As can be observed, the (rescaled) period of the oscillations in the medium and high density cases are similar; however the oscillations in the latter case are much more irregular.

\section{Conclusions}
\label{Sec:Conclusions}

In this work, we have considered a spacetime consisting of $N$ thin spherical concentric shells of matter separated by vacuum regions. By specifying an appropriate equation of state for this matter and based on the Israel junction formalism and the work in Refs.~\cite{Art:Brady1991Stability,Art:lemaitre2019equilibrium}, we have derived the relevant equilibrium conditions (see Eqs.~\eqref{Eq:equilibrium_one} and \eqref{Eq:equilibrium_two}) leading to static configurations and analyzed their stability with respect to small spherical perturbations of the shells. In the regime of a weak gravitational field and low temperature (i.e., the internal surface energy is much smaller than the rest one), the hydrostatic equilibrium conditions simplify considerably (see Eqs.~\eqref{Eq:shell_mass} and \eqref{Eq:hydrostatic_condition}) and can be interpreted in terms of Newtonian physics. Furthermore, we have shown that in this limit, the stability conditions are automatically satisfied.

In particular, we have shown that by choosing a configuration consisting of equidistant shells of equal mass, the low-temperature weak field limit can be maintained independently of the number of shells $N$. This allowed us to study the transmission properties of an outgoing monochromatic wave emanating from a source located at the center of the configuration and propagating through a configuration with an arbitrary number of shells. We solved the Regge-Wheeler equation perturbatively to first order in the small parameter $2m/r$ for a field with arbitrary spin $S$ and angular momentum number $\ell\geq S$ (see Eqs.~\eqref{Eq:Phi_ne} and \eqref{Eq:RadialW_sol}). To the best of our knowledge, the generalization to arbitrary values of $\ell$ is a new result. By using the transfer matrix formalism, we performed a comprehensive study of the transmission and reflection coefficients as a function of the frequency, the distance between the shells and the number of shells.

Our study revealed that the qualitative properties of the transmission and reflection coefficients depend fundamentally on the ratio between the areal radius of the first shell and the distance between the shells, which determines the average mass density of the system. While for low average density the reflection coefficient decays as $(\omega R_1)^{-4}$ and is almost independent of $N$, systems with high average density still decay in the same fashion but are modulated by wide oscillations in $\omega R_1$ which become more and more pronounced as $N$ increases. Most of these effects can be understood from the expression in Eq.~\eqref{Eq:Ref_coeff_Hfrequency} in the high-frequency limit.

In all cases we have studied, the reflection and transmission coefficients approach a finite value as $N$ becomes large. In particular, the transmission coefficient does not converge to zero in this limit. We attribute this to the fact that the transmission through the $j$-th shell is determined by its surface density $\Delta m/(4\pi R_j)^2$ (cf. Eqs.~\eqref{b_expansion}, \eqref{Eq:Transfer_matrix_T_E} and~\eqref{Eq:TExpl}). Since in our model $\Delta m$ is constant and $R_j$ grows linearly with $j$, the shells become progressively less dense and, therefore, more transparent which leads to the observed saturation as $N$ increases.

Future research could study the generalization of the present analysis to include even-parity linearized gravitational waves and different scenarios, like the propagation of a monochromatic plane wave from infinity through the spacetime region shaped by $N$ thin spherical concentric shells. Furthermore, it would be interesting to consider models with random fluctuations of the masses of the shells and the distances between them. Specifically, one could analyze in such a model the transmission properties of outgoing waves, and investigate whether disorder can produce localization effects~\cite{Art:Rossum1999multiple, book:ishimaru2017electromagnetic, book:sheng2006introduction, book:markos2008wave}.

\subsection*{Acknowledgements}

We thank Francisco Astorga, Alberto Diez-Tejedor, Ulises Nucamendi, and Emilio Tejeda for fruitful discussions.  R.O.A.C. was supported by a CONAHCyT doctoral scholarship. This work was partially supported by CONAHCyT Projects No. CF 2019/376127 and CBF-2023-2024-3116 and by CIC grants No. 18090 and 18315 of the Universidad Michoacana de San Nicol\'as de Hidalgo.

\appendix

\section{Hydrostatic equilibrium for thin Newtonian shells}
\label{App:New_shells}

In this appendix we show that Eq.~\eqref{Eq:hydrostatic_condition} which was obtained in Sec.~\ref{SubSec:WeakField} by taking the nonrelativistic limit of the Israel junction conditions can also be directly derived from the condition of hydrostatic equilibrium of a thin shell in a purely Newtonian setting. Our arguments are based on Secs.~II A and B  of~\cite{Art:lemaitre2019equilibrium} and generalize the results of that work to the case in which the shell surrounds a spherical concentric mass distribution.

Hence, we consider a thin spherical shell of matter of mass $m_s$ and radius $R$ which is subject to two forces: the force due to the push of matter in the outward radial direction, generated by the surface pressure $p$, and the gravitational force which is directed towards the center of the configuration. Let us denote the radial components of these forces by $F_p$ and $F_g$, respectively, such that the total force acting on the shell is
\begin{equation}
F = F_p + F_g.
\label{Eq:newtonF_shell}
\end{equation}
The force $F_p$ is obtained by analyzing the work done by the surface pressure due to an increase $dA$ of the area of the shell:
\begin{equation}
	dW = p\cdot dA = p\cdot 8\pi R dR,
\end{equation}
resulting in the net radial force being
\begin{equation}
F_p = 8\pi p R.
\end{equation}
\begin{figure}[h!]
	\includegraphics[scale=0.75]{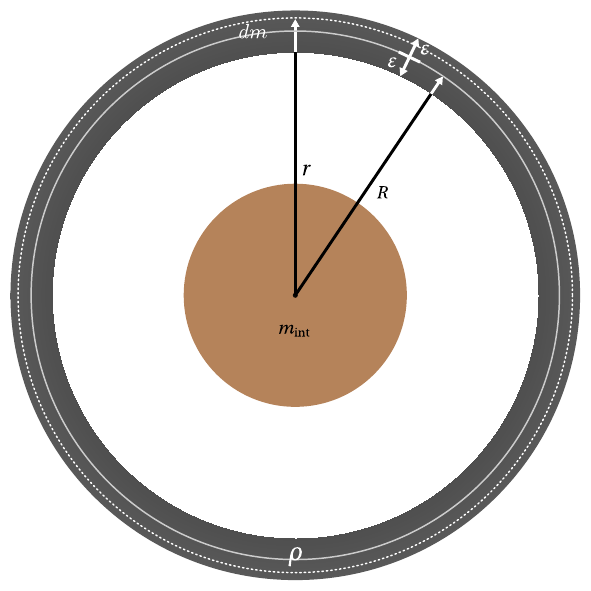}
	\caption{Configuration consisting of a spherical shell of width $2\varepsilon$ with a central object of mass $m_{\mathrm{int}}$ in its center.}
	\label{Fig:newton_shell}
\end{figure}
To calculate the gravitational force, it is convenient to perform the calculation starting from a shell of finite width $2\varepsilon$, as illustrated in Fig.~\ref{Fig:newton_shell}, and taking the limit $\varepsilon\to 0$ at the end of the calculation. We assume that the shell has the following mass density profile
\begin{equation}
\rho(r) :=
	\begin{cases}
		\frac{\sigma}{2\varepsilon}, & R-\varepsilon < r < R + \varepsilon,\\
		0, & \text{otherwise},
	\end{cases}
\end{equation}
where $\sigma$ denotes the surface density in the limit $\varepsilon\to 0$. Accordingly, the total mass contained in a sphere of radius $r$ lying in the interval $R - \varepsilon < r < R + \varepsilon$ is
\begin{equation}
m(r) = m_{\mathrm{int}} + 4\pi\sigma\frac{r^3 - (R-\varepsilon)^3}{6\varepsilon},
\end{equation}
with $m_{\mathrm{int}}$ the mass of the central object. Note that the mass of the shell, $m_s = m(R+\varepsilon) - m(R-\varepsilon)$, converges to $4\pi R^2\sigma$ in the limit $\varepsilon\to 0$, as required. The gravitational force acting on the layer of radius $r$ is
\begin{equation}
dF_g = -\frac{G m(r) dm(r)}{r^2},\qquad
dm(r) = 4\pi r^2\rho(r) dr.
\end{equation}
Hence, the total gravitational force acting on the shell is
\begin{eqnarray}
F_g &=&  -4\pi G\int\limits_{R-\varepsilon}^{R+\varepsilon} m(r) \rho(r) dr
\nonumber\\
 &=& -4\pi G\sigma\left[ m_{\text{int}} + \frac{2\pi\sigma}{3}(3R^2 - 2R\varepsilon + \varepsilon^2) \right].
\end{eqnarray}
Taking the limit $\varepsilon\to 0$ we obtain for the total force
\begin{equation}
F = 8\pi p R - 2\pi G\sigma\left( 2m_{\text{int}} + m_s \right),
\end{equation}
and hence the condition for hydrostatic balance is
\begin{equation}
2m_{\mathrm{int}} + m_s = \frac{4R}{G}\frac{p}{\sigma}.
\end{equation}
To apply this result to the $j$-th shell of the configurations in Sec.~\ref{Sec:Spacetime} we substitute $m_{\mathrm{int}}\mapsto m_{j-1}$, $m_s\mapsto m_j - m_{j-1}$, $R\mapsto R_j$, $p\mapsto p_j$, $\sigma\mapsto \sigma_{\rest,j}$ and use geometrized units $G=1$. This leads to Eq.~(\ref{Eq:hydrostatic_condition}) as was to be shown.

\section{Error estimate for the linear approximation of finite matrix products}
\label{App:TranM_T_error}

In this appendix we consider a finite product of matrices of the form
\begin{equation}
\mathbb{M}_N := \left( \mathds{1} + \epsilon_{N}\mathbb{B}_{N} \right)
\left( \mathds{1} + \epsilon_{N-1}\mathbb{B}_{N-1} \right)\cdots
\left( \mathds{1} + \epsilon_{1}\mathbb{B}_1 \right),
\label{eq:prod_nearby_mat}
\end{equation}
with a decreasing sequence $\epsilon := \epsilon_1 > \epsilon_2 > \ldots > \epsilon_{N-1} > \epsilon_N > 0$ of positive small numbers and uniformly bounded matrices $\mathbb{B}_j$, i.e., there is a constant $b$ such that
\begin{equation}
\| \mathbb{B}_k \| \leq b
\label{Eq:BkBound}
\end{equation}
for all $k\geq 1$. Consider the first-order approximation
\begin{equation}
\mathbb{M}_N^{(1)} := \mathds{1} + \sum\limits_{j=1}^N \epsilon_j\mathbb{B}_j,
\end{equation}
such that $\mathbb{M}_N = \mathbb{M}_N^{(1)} + {\cal O}(\epsilon^2)$. In general the quadratic error term ${\cal O}(\epsilon^2)$ depends on $N$. However, we show in the following that if the sequence $\epsilon_k$ falls off sufficiently fast, then the error term has an upper bound proportional to $\epsilon^2$  that is independent of $N$.

To analyze this, we note that the residual $R_N := \mathbb{M}_N - \mathbb{M}_N^{(1)}$ can be written as
\begin{eqnarray}
R_N &=& \sum\limits_{k_1 > k_2} \epsilon_{k_1}\epsilon_{k_2} \mathbb{B}_{k_1}\mathbb{B}_{k_2}
 + \sum\limits_{k_1 > k_2 > k_3} \epsilon_{k_1}\epsilon_{k_2}\epsilon_{k_3} \mathbb{B}_{k_1}\mathbb{B}_{k_2}\mathbb{B}_{k_3}
\nonumber\\
 &+& \ldots + \epsilon_N\epsilon_{N-1}\cdots\epsilon_1\mathbb{B}_{N}\mathbb{B}_{N-1}\cdots\mathbb{B}_{1}.
\label{Eq:Rn}
\end{eqnarray}
Taking into account Eq.~(\ref{Eq:BkBound}) and defining $s_N := \sum\limits_{k=1}^N \epsilon_k$ one finds
\begin{eqnarray}
\| R_N \| &\leq& b^2\sum\limits_{k_1 > k_2} \epsilon_{k_1}\epsilon_{k_2} 
 + b^3\sum\limits_{k_1 > k_2 > k_3} \epsilon_{k_1}\epsilon_{k_2}\epsilon_{k_3}
\nonumber\\
 &+& \ldots + b^N\epsilon_N\epsilon_{N-1}\cdots\epsilon_1
\nonumber\\
 &\leq& \frac{b^2}{2} s_N^2 + \frac{b^3}{3!} s_N^3 + \ldots + \frac{b^N}{N!} s_N^N
\nonumber\\
 &\leq& h(b s_N),
\end{eqnarray}
with the function $h(x) := e^x - 1 - x$. Using the estimate $h(x) \leq \frac{1}{2} x^2 e^x$ one concludes that
\begin{equation}
\| R_N \| \leq \frac{1}{2} (b s_N)^2 e^{b s_N},\qquad
s_N := \sum\limits_{k=1}^N \epsilon_k,
\end{equation}
for all $N\in \Natural$. In particular, if
\begin{equation}
s_\infty := \sum\limits_{k=1}^\infty \epsilon_k < \infty
\end{equation}
converges, it follows that $R_N$ has an upper bound that is independent of $N$. For the calculations in Sec.~\ref{Sec:TransferMatrix}, the relevant model is $\epsilon_k = \epsilon/k^2$; therefore $s_\infty = \epsilon\times\pi^2/6$ and one obtains the uniform bound
\begin{equation}
\| R_N \| \leq \frac{\pi^4 b^2}{36}\epsilon^2,
\end{equation}
for all $N\in \Natural$ and small enough $\epsilon > 0$ such that $\pi^2 b\epsilon\leq 6\log(2)$.

If $\epsilon_k$ is small but is not summable one cannot expect such a uniform bound in general. For example, if all $\mathbb{B}_k$'s are equal to the identity matrix and $\epsilon_k = \epsilon/k$, then it follows from Eq.~(\ref{Eq:Rn}) that
\begin{eqnarray}
\| R_N \| &\geq& 1 + \sum\limits_{k_1 > k_2} \epsilon_{k_1}\epsilon_{k_2} 
\nonumber\\
 &=& 1 + \frac{\epsilon^2}{2}\left[ \left( \sum\limits_{k=1}^N \frac{1}{k} \right)^2
 - \sum\limits_{k=1}^N \frac{1}{k^2} \right],
\nonumber\\
 &\geq& 1 + \frac{\epsilon^2}{2}\left[ \left( \sum\limits_{k=1}^N \frac{1}{k} \right)^2
 - \frac{\pi^2}{6} \right],
\end{eqnarray}
which diverges like $(\log N)^2$ when $N\to \infty$ even if $\epsilon$ is small.

\section{Electromagnetic flux through a sphere}
\label{App:EM_flux}

In this appendix we show that the electromagnetic flux through a sphere computed from the electromagnetic energy-momentum tensor is, up to a factor, equal to the expression obtained from the effective energy-momentum tensor with $S=1$ in Sec.~\ref{Sec:TR}.

Recall that in terms of the electromagnetic field tensor $F_{\mu\nu}$ the electromagnetic energy-momentum tensor $T_{em}^{\mu\nu}$ is given by
\begin{equation}
(T_{em})^\mu{}_\nu = F^{\mu\alpha}F_{\nu\alpha} - \frac{1}{4}\delta^\mu{}_\nu F_{\alpha\beta}F^{\alpha\beta},
\end{equation}
and recall that only its $rt$-component is needed to compute the average energy flow per unit time from Eq.~\eqref{Eq:flux_PUT}. In particular, the radial component of the electromagnetic current is
\begin{equation}
J^r_{(\mathrm{em})} = -(T_{em})^r{_t} = -F^{rA}F_{tA},
\label{Eq:Jr}
\end{equation}
where $A = \vartheta,\varphi$ refer to the spherical components. In the following, we express this quantity in terms of the functions $\Phi_{\ell m}$ that appear in Eq.~\eqref{Eq:PsiExpansion}.

To this purpose, it is necessary to know the relation between the components $F_{tA}$ and $F_{rA}$ of the electromagnetic field tensor and the functions $\Phi_{\ell m}$. From Sec.~III B in Ref.~\cite{Art:linearpert2013Chaverra} one can deduce the relation between $F_{\mu\nu}$ and the master scalars $\Phi^{(\pm)}$ in the even- and odd-parity sectors (denoted by $\Phi$ and $\nu^{(inv)}$ in ~\cite{Art:linearpert2013Chaverra}), which are related to our functions $\Phi_\ell(t,r) = \Phi^{(\pm)}_{\ell m}(t,r)$ satisfying the RW equation~\eqref{Eq:ReggeWheeler} with $S=1$ according to
\begin{equation}
\Phi^{(\pm)}(t,r,\vartheta,\varphi)
  = \sum_{\ell=1}^{\infty}\sum_{m=-\ell}^{\ell}\Phi^{(\pm)}_{\ell m}(t,r)Y^{\ell m}(\vartheta,\varphi).
\label{Eq:Phipm_Exp}
\end{equation}
In terms of these master scalars one has
\begin{align}
F_{tA} & = -\hat{\nabla}_{A}\hat{\Delta}^{-1}\left( \tilde{\varepsilon}^{a}{_t} \partial_a\Phi^{(+)} \right)
 + \hat{\varepsilon}_A{^B}\hat{\nabla}_{B}\partial_t\Phi^{(-)} ,
 \label{Eq:FtA}\\
F_{rA} & =
	- \hat{\nabla}_A \hat{\Delta}^{-1}\left( \tilde{\varepsilon}^a{_r}\partial_a \Phi^{(+)} \right)
	+ \hat{\varepsilon}_A{^B}\hat{\nabla}_B\partial_r \Phi^{(-)} ,
\label{Eq:FrA}
\end{align}
where $\hat{\nabla}_A$, $\hat{\Delta}$, $\hat{\varepsilon}_{AB}$ are the covariant derivative, the Laplacian, and the volume form associated with the unit two-sphere $S^2$, respectively, and $\tilde{\varepsilon}_{ab}$ is the volume form with respect to the two-metric $\tilde{g}_{ab} dx^a dx ^b = -\mathcal{N}(r) dt^2 + dr^2/\mathcal{N}(r)$, $\mathcal{N}(r) := 1 - 2m(r)/r$.

Using the fact that $\tilde{\varepsilon}_{tr} = -\tilde{\varepsilon}_{rt} = 1$ and substituting Eqs.~(\ref{Eq:FtA},\ref{Eq:FrA}) into Eq.~(\ref{Eq:Jr}) yields, after integrating and using integration by parts,
\begin{equation}
\int\limits_{S_{r}^{2}}d\Omega r^2 J_{(\mathrm{em})}^{r} 
 = \sum\limits_{\pm}\int\limits_{S_{r}^{2}}d\Omega
\left( \frac{\partial \Phi^{(\pm)}}{\partial r_*} \right) \hat{\Delta}^{\mp 1}\left( \frac{\partial \Phi^{(\pm)}}{\partial t} \right).
\end{equation}
Introducing the expansion~(\ref{Eq:Phipm_Exp}) into the above and using the orthonormality property of the spherical harmonics yields
\begin{align}
&\int\limits_{S_{r}^{2}}d\Omega r^2 J_{(\mathrm{em})}^{r}
\nonumber\\
 &= -\re\sum\limits_{\ell m \pm}
 [\ell(\ell+1)]^{\mp 1} \left( \overline{\frac{\partial \Phi_{\ell m}^{(\pm)}}{\partial r_*}} \right)\left( \frac{\partial \Phi_{\ell m}^{(\pm)}}{\partial t} \right).
\end{align}
Up to the factors $[\ell(\ell+1)]^{\mp 1}$ (which can be absorbed into the definitions of $\Phi_{\ell m}^{(\pm)}$) this agrees precisely with the expression on the first line of Eq.~(\ref{twoterms}). Therefore, for $S=1$, we obtain expressions for the electromagnetic flux which are equivalent to the ones derived in Sec.~\ref{flux_sphere}. In particular, the reflection and transmission coefficients defined in Eq.~(\ref{Eq:TR}) yield identical results in both cases.

\section{Properties of auxiliary matrices}
\label{App:Aux_M}

In this appendix it will be shown that the diagonal elements of the matrix $\mathbb{B}_\ell(R_j)$ defined in Eq.~\eqref{Eq:Matrix_B_ell} are purely imaginary. This is based on the fact that the parameter $2m_j/R_j$ is small and it is known that the elements of the matrix $\mathbb{D}_\ell(R_j)$ satisfy
\begin{align}
\left[ \mathbb{D}_\ell(R_j) \right]_{12} &= \overline{\left[ \mathbb{D}_\ell(R_j) \right]_{11}}, 
\label{Eq:D12}\\
\left[ \mathbb{D}_\ell(R_j) \right]_{22} &= \overline{\left[ \mathbb{D}_\ell(R_j) \right]_{21}},
\label{Eq:D22}
\end{align}
as follows from the definition in Eq.~\eqref{Eq:Matrix_D_ell_O} and the symmetry~(\ref{Eq:XnenwSym}) between the in- and outgoing solutions. Of course, the properties~(\ref{Eq:D12},\ref{Eq:D22}) also apply to the matrices $\mathbb{D}_\ell(R_j)^{(0)}$ and $\mathbb{D}_\ell(R_j)^{(1)}$ in the sum~\eqref{d_split}. Furthermore, recall that the matrices $\mathbb{D}_\ell(R_j)$ and $\mathbb{D}_\ell(R_j)^{(0)}$ have determinant $2s = -2i\omega$. As a consequence, one can verify that the matrix elements of $\left[\mathbb{D}_\ell(R_j)^{(0)}\right]^{-1}$ satisfy
\begin{align}
	\left[\mathbb{D}_\ell(R_j)^{(0)}\right]^{-1}_{21} &=
	 \overline{\left[\mathbb{D}_\ell(R_j)^{(0)}\right]^{-1}_{11}}, \\
	\left[\mathbb{D}_\ell(R_j)^{(0)}\right]^{-1}_{22} &=
	\overline{\left[\mathbb{D}_\ell(R_j)^{(0)}\right]^{-1}_{12}},
\end{align}
which implies that
\begin{align}
\left[\mathbb{B}_\ell(R_j)\right]_{22} = \overline{\left[\mathbb{B}_\ell(R_j)\right]_{11}}, 
\label{Eq:B22B11}\\
\left[\mathbb{B}_\ell(R_j)\right]_{21} = \overline{\left[\mathbb{B}_\ell(R_j)\right]_{12}}.
\end{align}
Finally, applying the determinant to both sides of Eq.~\eqref{Eq:Matrix_D_ell} and using the aforementioned properties one finds
\begin{align}
	1 &= \det\left(\mathds{1} + \frac{2m_j}{\omega R_j^2} \mathbb{B}_\ell(R_j)\right) \\
	&= 1 + \frac{2m_j}{\omega R_j^2}\tr \mathbb{B}_\ell(R_j) + \mathcal{O}\left(\left( \frac{2m_j}{\omega R_j^2}\right)^2\right),
\end{align}
showing that the trace of $\mathbb{B}_\ell(R_j)$ is zero. Together with Eq.~\eqref{Eq:B22B11} this yields
\begin{equation}
\left[\mathbb{B}_\ell(R_j)\right]_{11} +  \overline{\left[\mathbb{B}_\ell(R_j)\right]_{11}} = 0,
\end{equation}
which implies that the diagonal elements of $\mathbb{B}_\ell(R_j)$ are purely imaginary numbers.

\section{Independent WKB calculation for the transfer matrix}
\label{App:WKB}

In this appendix we perform an independent calculation for the transfer matrix in the high-frequency limit which is based on the WKB approximation. The calculation does not make use of the expansion of the RW equation in $2M/r$, and hence it provides an independent verification for the results obtained in Sec.~\ref{Sec:TransferMatrix} for high frequencies.

We start by introducing the ansatz $\Phi_\ell(t,r) = e^{-i\omega t} u(r)$ into the RW equation~\eqref{Eq:ReggeWheeler}, where from now on we drop the index $\ell$ for notational simplicity. This yields the time-independent Schr\"odinger equation
\begin{equation}
\left[ - \frac{\partial^{2}}{\partial r_{*}^{2}} + V(r) \right] u(r) = E u(r),
\label{Eq:RWTISE}
\end{equation}
with $E = \omega^2$ and the potential
\begin{equation}
V(r) := \mathcal{N}(r)\left[ \frac{\ell(\ell+1)}{r^2} + (1-S^{2})\frac{2m(r)}{r^3} \right].
\label{Eq:V}
\end{equation}
As in Appendix~\ref{App:EM_flux}, $\mathcal{N}(r)$ refers to the function $\mathcal{N}(r) := 1 - 2m(r)/r$, and we also recall that the relation between the tortoise coordinate $r_*$ and the areal radial coordinate $r$ is given by Eq.~\eqref{Eq:tortoise_coordinate} or~\eqref{Eq:explicit_tortoise_coo}. As in Sec.~\ref{Sec:RW} we first compute the solution in the region between two shells, where $m$ is constant and the potential $V(r)$ is smooth and then match the solutions across the shells using the matching conditions derived in Sec.~\ref{SubSec:WaveFunctionMatching}. 

In the WKB approximation (see for instance chapters 6 and 7 in Ref.~\cite{Book:Olver1974Introduction}) the in- and outgoing solutions of Eq.~\eqref{Eq:RWTISE} in the region $R_j < r < R_{j+1}$ are given by
\begin{equation}
u_{\nwarrow,j}(r) := \frac{e^{-i\phi_j(r)}}{\sqrt{k_j(r)}},\quad
u_{\nearrow,j}(r) := \frac{e^{i\phi_j(r)}}{\sqrt{k_j(r)}},
\label{Eq:WKBInOut}
\end{equation}
with the phase function
\begin{equation}
\phi_j(r) := \int\limits^r k_j(s) ds_*.
\end{equation}
Here, $k_j(r) := \sqrt{E - V_j(r)}$ with $V_j$ being the potential $V$ in the region $R_j < r < R_{j+1}$ where $m(r) = m_j$ is constant, and $ds_* = ds/\mathcal{N}_j(s)$. The approximation is valid as long as
\begin{equation}
\left| \frac{\mathcal{N}_j(r) k_j'(r)}{k_j(r)^2} \right| 
 = \frac{1}{2}\frac{| \mathcal{N}_j(r) V_j'(r)|}{|E - V_j(r)|^{3/2}} \ll 1,
\label{Eq:WKBCondition}
\end{equation}
where the prime denotes differentiation with respect to $r$. In particular, the condition~\eqref{Eq:WKBCondition} is satisfied for $E = \omega^2 \gg V(r)$.

In analogy with Eq.~\eqref{Eq:MonochromaticInOut} the WKB solution in the region $R_j < r < R_{j+1}$ is
\begin{equation}
u_j(r) = \Lambda_j u_{\nwarrow,j}(r) + \Upsilon_j u_{\nearrow,j}(r),
\end{equation}
and the matching conditions can be written in the same form as Eq.~\eqref{Eq:System_bound_cond}, that is
\begin{equation}
\mathbb{D}_j(R_{j})\left(\begin{array}{c}
\Upsilon_{j}\\
\Lambda_{j}
\end{array}\right) = 
\mathbb{D}_{j-1}(R_{j})\left(\begin{array}{c}
\Upsilon_{j-1}\\
\Lambda_{j-1}
\end{array}\right),
\label{Eq:System_bound_condWKB}
\end{equation}
with
\begin{equation}
\mathbb{D}_{j}(r) := 
\left(\begin{array}{rr}
u_{\nearrow,j}(r) & u_{\nwarrow,j}(r) \\
\mathcal{N}_j(r) u'_{\nearrow,j}(r) & \mathcal{N}_j(r) u'_{\nwarrow,j}(r)
\end{array}\right).
\label{Eq:Matrix_D_ell_OWKB}
\end{equation}
Using Eq.~\eqref{Eq:WKBInOut} one finds
\begin{equation}
\mathcal{N}_j(r) u'_{\nearrow,j}(r) = \left[ ik_j(r) - \frac{1}{2}\frac{\mathcal{N}_j(r)k_j'(r)}{k_j(r)} \right]
u_{\nearrow,j}(r),
\end{equation}
and similarly for the derivative of $u_{\nwarrow,j}(r)$. Owing to condition~\eqref{Eq:WKBCondition} one can neglect the second term on the right-hand side, which leads to the following simplification:
\begin{equation}
\mathbb{D}_{j}(r) = 
\left(\begin{array}{rr}
u_{\nearrow,j}(r) & u_{\nwarrow,j}(r) \\
i k_j(r) u_{\nearrow,j}(r) & - i k_j(r) u_{\nwarrow,j}(r) 
\end{array}\right).
\label{Eq:Matrix_D_ell_OWKB2}
\end{equation}
Note that $\det\left( \mathbb{D}_{j}(r) \right) = -2i$. From this, one finds
\begin{equation}
\left(\begin{array}{c}
\Upsilon_{j}\\
\Lambda_{j}
\end{array}\right) = \mathbb{M}_j\left(\begin{array}{c}
\Upsilon_{j-1}\\
\Lambda_{j-1}
\end{array}\right),
\end{equation}
with the WKB transfer matrix
\begin{equation}
\mathbb{M}_j := \left(\begin{array}{rr} 
\cosh(\Theta_j) e^{-i\Delta_j} & \sinh(\Theta_j) e^{-i\Sigma_j} \\
\sinh(\Theta_j) e^{i\Sigma_j} & \cosh(\Theta_j) e^{i\Delta_j} 
\end{array}\right)
\end{equation}
where $\Theta_j$, $\Delta_j$, and $\Sigma_j$ are defined as
\begin{eqnarray}
e^{\Theta_j} &:=& \sqrt{\frac{k_j(R_j)}{k_{j-1}(R_j)}},
\\
\Delta_j &:=& \phi_j(R_j) - \phi_{j-1}(R_j),
\\
\Sigma_j &:=& \phi_j(R_j) + \phi_{j-1}(R_j).
\end{eqnarray}

In the high-frequency limit $\omega^2 \gg V_j(r)$ we can expand
\begin{equation}
k_j(r) = \omega\sqrt{1 - \frac{V_j(r)}{\omega^2}}
 = \omega\left[ 1 - \frac{V_j(r)}{2\omega^2} 
  + {\cal O}\left( \frac{V_j(r)^2}{\omega^4} \right) \right],
\label{Eq:kjApprox}
\end{equation}
which yields
\begin{equation}
e^{\pm\Theta_j} = 1 \mp \frac{V_j(R_j) - V_{j-1}(R_j) }{4\omega^2}
 + {\cal O}\left( \frac{V_j^2}{\omega^4} \right).
\end{equation}
Using Eq.~\eqref{Eq:V}, recalling $m_j - m_{j-1} = \Delta m$ and abbreviating $\mu_j := (m_j + m_{j+1})/R_j$, one finds
\begin{eqnarray}
\cosh(\Theta_j) &=& 1 + {\cal O}\left( \frac{V_j^2}{\omega^4} \right),
\\
\sinh(\Theta_j) &=& \frac{\Delta m}{2R_j^3\omega^2}\left[ \ell(\ell+1) 
 - (1-S^2)( 1 - 2\mu_j) \right] 
\nonumber\\
 &+& {\cal O}\left( \frac{V_j^2}{\omega^4} \right).
\end{eqnarray}
Furthermore, after integrating Eq.~\eqref{Eq:kjApprox} and adjusting the integration constant, one obtains
\begin{equation}
\phi_j(r) = \omega r_* 
 + \frac{1}{2\omega}\left[ \frac{\ell(\ell+1)}{r} + \frac{(1-S^2)m_j}{r^2} \right]
 +  {\cal O}\left( \frac{V_j^2 R_j}{\omega^3} \right).
\end{equation}
This implies
\begin{eqnarray}
\Delta_j &=& \frac{1-S^2}{2\omega R_j}\frac{\Delta m}{R_j} + {\cal O}\left( \frac{V_j^2 R_j}{\omega^3} \right),\\
\Sigma_j &=& \left. 2\omega r_* \right|_{R_j} + \frac{1}{\omega R_j}\left[ \ell(\ell+1) + (1-S^2)\mu_j \right] 
\nonumber\\
 &+& {\cal O}\left( \frac{V_j^2 R_j}{\omega^3} \right).
\end{eqnarray}
Taking into account that $V_j$ is of the order $2m_j/R_j^3\leq 1/R_j^2$ when $\ell=0$ and of the order $1/R_j^2$ when $\ell \geq 1$, one obtains the following matrix elements:
\begin{eqnarray}
M_{j,11} &=& 1 - i \frac{1-S^2}{2\omega R_j}\frac{\Delta m}{R_j} 
 - \frac{1}{2}\left( \frac{1-S^2}{2\omega R_j}\frac{\Delta m}{R_j} \right)^2
\nonumber\\
 &+& {\cal O}\left( \frac{1}{\omega^3 R_j^3} \right),\\
M_{j,12} &=& \frac{\Delta m}{2\omega^2 R_j^3}
\left[ \ell(\ell+1) - (1-S^2)(1 - 2\mu_j) \right] e^{-2i\left. \omega r_* \right|_{R_j}} \nonumber\\
 &+& {\cal O}\left( \frac{1}{\omega^3 R_j^3} \right).
\end{eqnarray}
The terms that are zeroth and first order in $\Delta m/R_j$ agree precisely with the results obtained in the high-frequency limit in Sec.~\ref{Sec:PertTransferMatrix}, see Eqs.~(\ref{b_0},\ref{b_1},\ref{m_close_to_1}).

\bibliography{refs_Wprop}
\end{document}